\documentclass[sigconf]{acmart}

\AtBeginDocument{%
  }

\copyrightyear{2024}
\acmYear{2024}
\setcopyright{rightsretained}
\acmConference[CHI '24]{Proceedings of the CHI Conference on Human Factors in Computing Systems}{May 11--16, 2024}{Honolulu, HI, USA}
\acmBooktitle{Proceedings of the CHI Conference on Human Factors in Computing Systems (CHI '24), May 11--16, 2024, Honolulu, HI, USA}
\acmDOI{10.1145/3613904.3642631}
\acmISBN{979-8-4007-0330-0/24/05}

\usepackage{subcaption}

\usepackage{multirow}

\usepackage{array}
\newcolumntype{P}[1]{>{\centering\arraybackslash}p{#1}}
\newcolumntype{M}[1]{>{\centering\arraybackslash}m{#1}}

\usepackage{cleveref}
\usepackage{hyperref}

\definecolor{pal1}{HTML}{332288}
\definecolor{pal2}{HTML}{117733}
\definecolor{pal3}{HTML}{44AA99}
\definecolor{pal4}{HTML}{88CCEE}
\definecolor{pal5}{HTML}{DDCC77}
\definecolor{pal6}{HTML}{CC6677}
\definecolor{pal7}{HTML}{AA4499}
\definecolor{pal8}{HTML}{882255}

\begin{document}

\title[The Mixed Reality Concerns (MRC) Questionnaire]{Assessing User Apprehensions About Mixed Reality Artifacts and Applications: The Mixed Reality Concerns (MRC) Questionnaire}

\author{Christopher Katins}
\email{christopher.katins@hu-berlin.de}
\affiliation{%
  \institution{HU Berlin}
  \city{Berlin}
  \country{Germany}
}
\orcid{0000-0001-6257-7057}

\author{Paweł W. Woźniak}
\email{pawel.wozniak@chalmers.se}
\affiliation{%
  \institution{Chalmers University of Technology}
  \city{Gothenburg}
  \country{Sweden}
}
\orcid{0000-0003-3670-1813}

\author{Aodi Chen}
\email{aodi.chen@student.hu-berlin.de}
\affiliation{%
  \institution{HU Berlin}
  \city{Berlin}
  \country{Germany}
}
\orcid{0009-0005-0702-6726}

\author{Ihsan Tumay}
\email{ihsan.tumay@hu-berlin.de}
\affiliation{%
  \institution{HU Berlin}
  \city{Berlin}
  \country{Germany}
}
\orcid{0009-0008-7869-8505}

\author{Luu Viet Trinh Le}
\email{leluvitr@hu-berlin.de}
\affiliation{%
  \institution{HU Berlin}
  \city{Berlin}
  \country{Germany}
}
\orcid{0009-0002-1038-600X}

\author{John Uschold}
\email{john.uschold@student.hu-berlin.de}
\affiliation{%
  \institution{HU Berlin}
  \city{Berlin}
  \country{Germany}
}
\orcid{0009-0001-4460-088X}

\author{Thomas Kosch}
\email{thomas.kosch@hu-berlin.de}
\affiliation{%
  \institution{HU Berlin}
  \city{Berlin}
  \country{Germany}
}
\orcid{0000-0001-6300-9035}

\renewcommand{\shortauthors}{Katins et al.}

\begin{abstract}
Current research in Mixed Reality (MR) presents a wide range of novel use cases for blending virtual elements with the real world. This yet-to-be-ubiquitous technology challenges how users currently work and interact with digital content. While offering many potential advantages, MR technologies introduce new security, safety, and privacy challenges. Thus, it is relevant to understand users' apprehensions towards MR technologies, ranging from security concerns to social acceptance. To address this challenge, we present the Mixed Reality Concerns (MRC) Questionnaire, designed to assess users' concerns towards MR artifacts and applications systematically. The development followed a structured process considering previous work, expert interviews, iterative refinements, and confirmatory tests to analytically validate the questionnaire. The MRC Questionnaire offers a new method of assessing users' critical opinions to compare and assess novel MR artifacts and applications regarding security, privacy, social implications, and trust.

\end{abstract}




\maketitle

\section{Introduction}
\label{sec:intro}
Mixed Reality (MR)~\cite{speicherWhatMixedReality2019} research is a growing field covering a broad spectrum of technologies and applications that blur the boundaries between digital and real worlds. Considering the evolution of MR over the past years, we observed that many innovations have primarily brought incremental improvements to MR technologies. As a consequence, MR devices become more commonly available through smartphones~\cite{leeDualMRInteractionMixed2018} and even more interwoven by using head-mounted displays~\cite{mehrfardComparativeAnalysisVirtual2019, kiyokawaIntroductionHeadMounted2007}. Fueled by the commercial success of the Microsoft HoloLens 2\footnote{\url{https://www.microsoft.com/en-us/hololens}, last accessed on 2023-12-12.} in industry settings and further expectations towards the Apple Vision Pro\footnote{\url{https://www.apple.com/apple-vision-pro}, last accessed on 2023-12-12.} in consumer use, MR might become omnipresent soon.

The transition from fiction to reality has steadily progressed in recent decades with the continued research in this field and the emergence of commercially available MR products. Previous research has extensively investigated use cases (e.g., in the context of work~\cite{moserMixedRealityApplications2019} or education~\cite{hughesMixedRealityEducation2005}) as these technologies become increasingly accessible. At the same time, evaluating their usability and potential benefits is essential, as well as understanding the concerns and apprehensions that MR devices raise with their integration into our lives. Existing issues related to hardware performance, software optimization, and interaction design tend to improve over time as computational power increases and hardware shrinks. As a result, technical challenges that currently hinder seamless MR experiences will likely diminish over time. Yet, it is essential to recognize that the evolution of MR is not solely a matter of technological advancement. The challenge lies in addressing individuals' potential apprehensiveness about the technology.

In this context, a new challenge emerges for HCI: Understanding how individuals apprehend novel MR systems regarding their perceived concerns about this technology. Numerous methodologies and tools have been developed for evaluating user experience (e.g., UEQ~\cite{schreppConstructionBenchmarkUser2017a}), usability (e.g., SUS~\cite{grierSystemUsabilityScale2013}), or acceptance (e.g., TAM~\cite{davisUserAcceptanceComputer1989}). At the same time, researchers have rarely investigated users' apprehensions and concerns regarding novel MR technologies besides usability measures. Thus, measuring user apprehensions and concerns remains a research gap.

This paper presents the Mixed Reality Concern (MRC) Questionnaire to address these challenges. The MRC enables an evaluation that extends beyond usability and other aspects of the new system, encompassing potential concerns and apprehensions. Our systematic approach to developing this scale was based on the guidelines by Boateng et al.~\cite{boatengBestPracticesDeveloping2018}. Initially, a conceptual model of potential concerns was formulated, drawing from relevant research in the field. This model comprises four primary categories, with 30 subcategories that cover a broad spectrum of potential user concerns. The model is shown in \Cref{table:init-cat}. Subsequently, an initial set of 120 items derived from this conceptual model was generated. These items were then refined through expert feedback and underwent an exploratory factor analysis, resulting in the final scale composed of 9 items. A comprehensive evaluation of this scale followed to ensure the validity of its results.
Finally, we anticipate the MRC to complement current usability metrics by acting as a tool for researchers and practitioners to measure concerns towards their MR applications and artifacts. 
Given that MR is a technology distinct from traditional user interfaces and devices that increasingly proliferates into home and work environments~\cite{speicherWhatMixedReality2019}, users may assume implicit or explicit concerns that significantly influence their interaction with these artifacts. The questionnaire is designed to concentrate specifically on MR-related user concerns, facilitating practitioners in quickly and comprehensively understanding potential apprehensions that could affect the overall user experience.
\section{Related Work}
\label{sec:related}
With the rapid advancements in MR technology, understanding users' apprehensions about MR technology is crucial for its successful integration into everyday life. MR has shown tremendous potential in various domains, but its widespread adoption is impeded by several challenges that need to be addressed for it to become a mainstream technology~\cite{hughesMixedRealityEducation2005}. By giving an overview of current research about novel challenges in MR, we aim to provide a comprehensive backdrop against which user concerns can be effectively evaluated in the later sections. 

\subsection{Social Acceptance and Social Implications: Challenges to the Ubiquity of MR}
\label{sec:social-acc}
One of the critical barriers to the widespread acceptance of MR is the lack of social acceptance. A 2021 study by Thomas et al.~\cite{thomasFunctionalityAllThat2021} sheds light on the barriers to social acceptance surrounding MR devices. Despite the functional benefits of MR technology, the study reveals that certain facts genuinely worry everyday users. One of the primary barriers is the perceived social awkwardness associated with wearing MR devices in public, which can lead to feelings of self-consciousness and reluctance to embrace this technology. Moreover, the study mentions that the appearance and design of MR devices are critical factors influencing social acceptance, as aesthetically unappealing or intrusive devices may deter individuals from incorporating them into their daily lives. To foster broader social acceptance of MR, the study emphasizes the importance of improving the functionality and user experience and addressing these social and psychological barriers to ensure MR devices become seamlessly integrated into society's fabric.

Slater et al.~\cite{slaterEthicsRealismVirtual2020} determined a number of ethical considerations that ought to be considered in the future development of MR technologies. Next to common privacy concerns due to the vast amount of data collected by MR devices (further discussed in \Cref{sec:secsafpriv}), the publication illustrates how highly realistic VR and AR environments can impact users emotionally, psychologically, and socially. These impacts include but are not limited to the ubiquity of MR, akin to mobile technology, as it can impede meaningful real-world interactions, potentially resulting in social isolation. This shift towards MR may also cultivate a preference for virtual interactions over real-life ones, leading to societal withdrawal. Moreover, the potential ``superrealism'' of MR experiences may lead some individuals to neglect their physical well-being, paralleling extreme cases of excessive video game usage where the boundary between the virtual and physical worlds blurs. Immersive MR environments can also encourage imitative behaviors that individuals would typically avoid in reality, either through gradual exposure or emulation of actions taken by virtual characters. The persuasive power of MR, particularly in highly realistic iterations, raises ethical concerns when employed to modify emotions and behaviors for potentially harmful ends. Furthermore, this capacity to manipulate sensory experiences raises questions about the reliability of sensory evidence in both legal and societal contexts.

\subsection{Security, Safety, and Privacy: Common Threats in a New Environment}
\label{sec:secsafpriv}
According to Gugenheimer et al.~\cite{gugenheimerNovelChallengesSafety2022}, while a significant portion of research focuses on technological advancements in MR, it is equally crucial to emphasize research into the potential hazards and challenges that accompany these innovations.
They determined the well-established topics of security, safety, and privacy in general computer science to be relevant for the MR research. These aspects become more important since they proliferate into other research areas for wider adoption, including MR support at production lines~\cite{buettner2017design,nowak2020what}, education~\cite{knierim2020demonstrating,feger2022electronicsar}, or transportation~\cite{koschNotiBikeAssessingTarget2022,10.1145/3544549.3585742,10.1145/3626705.3627785} while changing the perception and interaction capacities of users~\cite{speicherWhatMixedReality2019, 10.1145/3411764.3445349, schoen2023tailor}. With such growth in MR, privacy concerns encompass two main viewpoints: that of the user and that of bystanders. User-related privacy issues revolve around the risks associated with biometric identification or surveillance of behavior and attention. In contrast, bystander privacy concerns how MR sensors, such as cameras, may impact individuals who did not consent to be observed by the technology~\cite{10339660}.

In the context of trust, Jian et al.~\cite{jianFoundationsEmpiricallyDetermined2000} discuss the increasing prevalence of automation in complex systems and everyday life. The authors review existing research on measuring trust in various contexts, such as social psychology and human-machine systems, highlighting the multidimensional nature of trust and the need for a more empirical understanding. Furthermore, the authors identify and scrutinize previous studies, including the lack of differentiation between trust and distrust, and emphasize the importance of assessing trust in the context of human-machine systems, leading to the necessity for the development of an empirically based tool for assessing trust in increasingly automated environments.

Further, Harborth and Pape ~\cite{harborthInvestigatingPrivacyConcerns2021} also report that technical assessments of risks related to MR reveal that the technology introduces new privacy concerns that require immediate attention. Individuals using MR genuinely worry about their privacy, and these apprehensions significantly deter technology adoption. The study highlights the importance of addressing these privacy risks promptly and effectively to foster trust and confidence among users.

A unique aspect emerging in MR research is ``immersive attacks,''~\cite{caseyImmersiveVirtualReality2021,azmandianHapticRetargetingDynamic2016,wilsonViolentVideoGames2018} which target users' physiological and psychological safety through perceptual manipulation rather than exploiting hardware or software vulnerabilities. These attacks leverage perceptional illusions and necessitate the development of protective layers to detect and prevent such manipulations, highlighting the distinctive challenges posed by MR technology.

Lastly, safety and health concerns are yet another barrier that must be addressed to facilitate the broader adoption of MR. Yuntao Guo et al.~\cite{guoSafetyHealthPerceptions2021} reported on the safety and health concerns associated with location-based MR gaming applications. As these games blur the lines between virtual and physical environments, potential risks and hazards emerge that can impact players' well-being. The study mentions that one primary concern is the distraction factor, where players may become engrossed in the game and fail to pay adequate attention to their surroundings, leading to accidents or injuries. Additionally, prolonged usage of MR gaming apps can result in physical strain, eye discomfort, and even musculoskeletal issues, especially when players engage in prolonged or repetitive gameplay~\cite{kaufeldOpticalSeethroughAugmented2022}. The study emphasizes the importance of understanding these safety and health implications, particularly for game developers and policymakers, to implement safety measures, provide user guidelines, and raise players' awareness of the responsible use of location-based MR gaming apps.

\subsection{Related Questionnaires}
\label{sec:rel-quest}
Next to the objective key challenges that pertain to MR, acquiring the feedback of users is invariably a crucial part of the development of new technologies, be it in the field of MR or elsewhere. To this end, numerous questionnaires and scales have been developed to assess various aspects of user experiences within this domain. However, it is essential to note that these existing questionnaires often focus on specific dimensions of user perceptions and do not comprehensively address the diverse spectrum of concerns that may arise. This section briefly overviews these related questionnaires, highlighting their strengths and limitations.

One of the most widely known measures of user acceptance of technology is the \textbf{Technology Acceptance Model (TAM)}, developed by Fred Davis in the 1980s~\cite{davisPerceivedUsefulnessPerceived1989,davisUserAcceptanceComputer1989}. It aims to measure acceptance by determining both the \textit{ease of use} and the \textit{perceived usefulness} of a technological system. The TAM has been further developed~\cite{venkateshTheoreticalExtensionTechnology2000,venkateshTechnologyAcceptanceModel2008}, and other publications aimed at extending the model by adding further factors, such as \textit{perceived enjoyment}~\cite{reseHowAugmentedReality2017,liawInvestigationUserAttitudes2003}.
Notable is also the \textbf{Attitudes toward Virtual Reality Technology Scale (AVRTS)}~\cite{bunzTAMAVRTSDevelopment2021}, using the TAM as an initial model to then further develop a scale to assess attitudes towards VR technologies.
All in all, the TAM, its variants, the AVRTS, and other commonly used scales in HCI research like the \textbf{System Usability Scale (SUS)}~\cite{grierSystemUsabilityScale2013}, the \textbf{AttrakDiff}~\cite{hassenzahlAttrakDiffFragebogenZur2003}, or the \textbf{User Experience Questionnaire (UEQ)}~\cite{schreppConstructionBenchmarkUser2017a} is based on assessing the acceptance, the general usability, the hedonic and pragmatic qualities, and the general user experience respectively.
While these scales excel at evaluating usability and gauging user affinity for a particular artifact, their design does not prioritize the measurement of concerns or unfavorable opinions regarding those devices.

The \textbf{Perceived Creepiness Technology Scale (PCTS)}~\cite{wozniakCreepyTechnologyWhat2021} stands out in this respect as it specifically seeks to evaluate an adverse emotion. The primary purpose of the \textbf{PCTS} is to allow designers and researchers to quickly assess new technologies that might elicit initial sensations of creepiness in users in that regard.
 
Next to the \textbf{AVRTS}, scales like the \textbf{Augmented Reality Immersion (ARI)} questionnaire~\cite{georgiouDevelopmentValidationARI2017} and various presence questionnaires~\cite{witmerMeasuringPresenceVirtual1998,regenbrechtMeasuringPresenceAugmented2021a} seek to ensure that the measurements are relevant and accurate when applied to MR use cases, necessitating the development of novel questionnaires tailored to these technologies.
The \textbf{Virtual reality sickness questionnaire (VRSQ)}~\cite{kimVirtualRealitySickness2018} and the \textbf{Augmented Reality Sickness Questionnaire (ARSQ)}~\cite{hussainAugmentedRealitySickness2023} aim to measure the immediate negative impact of MR on the users' well-being, but to the authors' best knowledge, no scales exist that aim to determine the long-term effect of MR on its users.

Remotely related is the Concerns-Based Adoption Model (CBAM) with its \textbf{Stages of Concern Questionnaire (SoCQ)}~\cite{georgeMeasuringImplementationSchools2008}, an educational framework developed in the late 1970s. It is designed to understand and facilitate the process of educational innovation and change, particularly in the context of school settings. 
Although the questionnaire may not be suitable for assessing concerns related to MR technology and its users, the stages it outlines provide valuable insights into how individuals perceive innovations and their potential reactions to them.

\begin{figure*}
    \centering
    \includegraphics[width=\textwidth]{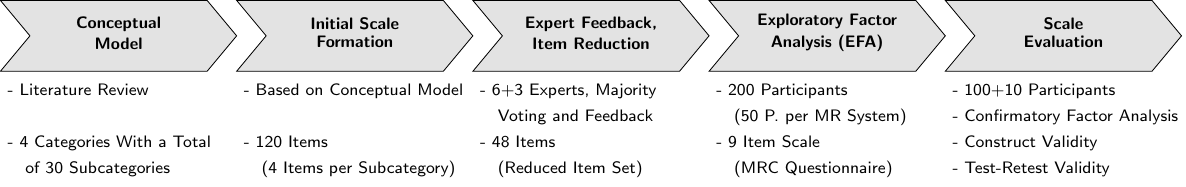}
    \caption{The process of developing the scale, as this paper outlines.}
    \label{fig:dev-process}
    \Description{The figure shows the linear progress of the development. There are five distinct steps: Conceptual Model (including a literature review), Initial Scale Formation (based on conceptual model), Expert Feedback and Item Reduction (with 6+3 experts, majority voting and feedback), Exploratory Factor Analysis (with 200 participants, 50 per MR system), and Scale Evaluation(with 100+10 participants, a confirmatory factor analysis, construct validity, and test-retest validity).}
\end{figure*}
\section{Conceptual Framework: Categorizing Concerns about MR}
\label{sec:concept}
Based on the findings of \Cref{sec:related}, a preliminary conceptual framework was developed to categorize potential user concerns about MR systems. As this classification is derived from related literature, it can logically only serve as a framework for classifying the ongoing research within this domain. Acknowledging that such categorizations may not always align with users' subjective concerns or considerations is essential. Hence, this only represents an initial basis from which the subsequent construction of the scale could proceed as further explained in \Cref{sec:formation}.

The decision to develop a preliminary conceptual framework for generating the questionnaire items rather than to base it on psychological models, such as the Innovation Resistance Theory (IRT)~\cite{ramConsumerResistanceInnovations1989}, was driven by the recognition that possible concerns regarding MR might extend beyond the generic barriers that are often defined for novel technologies or innovations as a whole. Herein, contemporary issues such as privacy, which are crucial in the field of MR, are often only implicitly addressed in existing models, if at all. Hence, deriving potential concerns from currently recognized challenges in MR was deemed more fitting, ensuring that the questionnaire reflects the nuanced research field of MR and addresses issues that may not be adequately captured by existing psychological models.

\begin{table}[ht]
    \centering
    \caption{The preliminary conceptual framework with its four categories and their respective subcategories aiming to classify potential user concerns regarding MR systems. This model will be used to develop the scale in the following.}
    \label{table:init-cat}
    \begin{tabular}{ M{.95\linewidth} }
        \hline 
        \noalign{\vskip 1mm}
        \textbf{\large User Concerns About MR Systems}\\
        \noalign{\vskip 1mm} 
        \hline 
        \noalign{\vskip 5mm} 
        \hline
        \noalign{\vskip .5mm}
        \textbf{Security~\cite{deguzmanSecurityPrivacyApproaches2019}} \\
        \noalign{\vskip .5mm}
        \hline
        \noalign{\vskip .5mm}
        \textit{Integrity} \\
        \textit{Non-Repudiation} \\
        \textit{Availability} \\
        \textit{Authorization} \\
        \textit{Authentification} \\
        \textit{Identification} \\
        \textit{Confidentiality} \\
        \noalign{\vskip .5mm}
        \hline
        \noalign{\vskip 5mm} 
        \hline
        \noalign{\vskip .5mm}
        \textbf{Privacy~\cite{deguzmanSecurityPrivacyApproaches2019}} \\
        \noalign{\vskip .5mm}
        \hline
        \noalign{\vskip .5mm}
        \textit{Anonymity \& Pseudonymity} \\
        \textit{Unlinkability} \\
        \textit{Unobservability \& Undetectability} \\
        \textit{Plausible Deniability} \\
        \textit{Content Awareness} \\
        \textit{Policy \& Consent Compliance} \\
        \noalign{\vskip .5mm}
        \hline 
        \noalign{\vskip 5mm} 
        \hline
        \noalign{\vskip .5mm}
        \textbf{Social Implications~\cite{slaterEthicsRealismVirtual2020}} \\
        \noalign{\vskip .5mm}
        \hline
        \noalign{\vskip .5mm}
        \textit{Social Isolation} \\
        \textit{Preference for Virtual Social Interactions} \\
        \textit{Body Neglect} \\
        \textit{Imitative Behavior} \\
        \textit{Persuasion} \\
        \textit{Unexpected Horror} \\
        \textit{Pornography and Exposure to Violence} \\
        \textit{Extreme Violence and Assault} \\
        \textit{Lack of Common Environments} \\
        \textit{Lack of Ground Truth} \\
        \textit{Persuasive Advertising} \\
        \noalign{\vskip .5mm}
        \hline 
        \noalign{\vskip 5mm} 
        \hline
        \noalign{\vskip .5mm}
        \textbf{Public Acceptance~\cite{guptaSociopsychologicalDeterminantsPublic2012}} \\
        \noalign{\vskip .5mm}
        \hline
        \noalign{\vskip .5mm}
        \textit{Perceived Health Implications} \\
        \textit{Social Outcast} \\
        \textit{Interactions} \\
        \textit{Trust} \\
        \textit{Family \& Friends} \\
        \textit{Perceived Risk} \\
        \noalign{\vskip .5mm}
        \hline 
    \end{tabular}
\end{table}

\subsection{Security and Privacy: Contrasting, yet not Mutually Exclusive}
\label{sec:secpriv}
The categorization of security threats in MR is based on the publication "Security and Privacy Approaches in Mixed Reality: A Literature Survey"~\cite{deguzmanSecurityPrivacyApproaches2019}. It compiles various strategies suggested to maintain the security and privacy of users and data within the realm of MR in previous work. Furthermore, the researchers combined the already existing security and privacy properties from previous work~\cite{howardSecurityDevelopmentLifecycle2006,kalloniatisAddressingPrivacyRequirements2008,dengPrivacyThreatAnalysis2011} for a final scale consisting of six security-related properties and six privacy-related properties on each end, with one property being related to both. They observed that specific security attributes may be simultaneously perceived as potential privacy risks. They note that this underscores the variations in the emphasis placed on these attributes or prerequisites by different\break stakeholders.

This categorization provides a comprehensive overview of the security and privacy risks in MR that are presently recognized in research and actively addressed, conceivably also covering the concerns that users of MR systems might have in this regard. As a result, the aforementioned properties form two of the four principal categories within our framework.

\subsection{Social Implications: Psychological Safety, Health, and Social Impact}
Safety, specifically psychological safety, is another novel challenge in MR~\cite{gugenheimerNovelChallengesSafety2022}. In this context, the publication "The Ethics of Realism in Virtual and Augmented Reality"~\cite{slaterEthicsRealismVirtual2020} identified eleven potential psychological and social implications that should be considered in the future development of MR.
Given the extensive range of potential social impacts, achieving comprehensive coverage is unattainable.
Yet, to consider a broad range of potential psychological and social concerns, we chose to integrate each implication as a subcategory under the respective factor.

\subsection{Public Acceptance: Perception and Trust}
Numerous factors can potentially shape the public's willingness to embrace novel technologies. The publication "Socio-psychological determinants of public acceptance of technologies: A review"~\cite{guptaSociopsychologicalDeterminantsPublic2012} sought to explore the psychological factors that underlie the societal acceptance of emerging technologies and assembled a list of the most frequently employed determinants found in related research. We opted for choosing a subset of these determinants that seemed fitting for the application regarding MR technology, especially considering the findings of \Cref{sec:social-acc}. Herein, the primary emphasis centers on the perception of the technology rather than its actual properties and the level of trust in these systems.

\section{Scale Formation}
\label{sec:formation}
After establishing a related-work-based initial conceptual framework for categorizing potential user concerns about MR, the subsequent phase involved developing a questionnaire that covers the genuine apprehensions of users. We followed a systematic procedure to accomplish this, as illustrated in \Cref{fig:dev-process}. This procedure is based on the scale development best practices proposed by Boateng et al.~\cite{boatengBestPracticesDeveloping2018}. 
This approach closely aligns with the methodology employed for developing the PCTS~\cite{wozniakCreepyTechnologyWhat2021}, which also aims to capture critical sentiments regarding novel technologies.


\subsection{Item Generation}
The initial items were generated by two researchers, creating four items for each subcategory of the conceptual framework, resulting in a total of 120 items. As the related work~\cite{deguzmanSecurityPrivacyApproaches2019,slaterEthicsRealismVirtual2020,guptaSociopsychologicalDeterminantsPublic2012} gave definitions for each property/implication, we generated similar, albeit slightly different phrasing to allow for a more nuanced set of items in the end. Afterward, the authors discussed the items and revised items that sounded too similar.
        
\subsection{Expert Feedback}
Two rounds of expert feedback were carried out to reduce the substantial pool of initial items. In the first round, six experts were asked to give feedback on the initial set of items and indicate whether they considered each item essential for such a scale. The experts were researchers in the fields of privacy, security, VR/AR, and general HCI.

The reduced set of items was chosen through majority voting, meaning that only if at least three experts indicated an item to be essential, it was retained, and all other items were discarded. The remaining items were then discussed and improved upon by the researchers based on the initial feedback of the experts. This resulted in a final set of 48 items.

Subsequently, another round of expert feedback was gathered for a final iteration, specifically regarding the phrasing to ensure that all items are easily comprehendable and sufficiently distinct. The three experts involved in the second round differed from those who participated in the initial round. They are researchers in the fields of VR/AR, human augmentation, and general HCI, respectively. Two of the experts had previous experience in developing questionnaires, while one expert, although knowledgeable about the process, had not previously engaged in questionnaire development.
To maintain balance and minimize potentially leading questions, half of the items in the final set were reversed. This was done to ensure that overwhelmingly negative phrasing would not skew responses, reducing bias where possible.

\subsection{First Survey}
\label{sec:1stsurvey}
After developing the reduced set of initial scale items, based on related work and expert feedback, a participant study was executed to refine the item set further through exploratory factor analysis. 
In accordance with the sample size recommendation by Comrey~\cite{comreyFactoranalyticMethodsScale1988}, $n=200$ participants were recruited. 

\subsubsection{Participants}
Prolific\footnote{\url{https://www.prolific.co}, last accessed on 2023-12-12.} was used to recruit participants, ensuring a more representative sample of subjects than through institute mailing lists or similar approaches. The participation was entirely voluntary, and the option to withdraw from the survey was available throughout. Participants were compensated with \pounds1.50 upon completing the survey, corresponding to an average hourly reward of \pounds15.15. 
The survey was conducted entirely online and took approximately 10 minutes to complete. The average age of participants was roughly 40 years ($\bar{x} = 39.65$, $s = 12.94$), 50\% identifying as male, 50\% identifying as female, and all either currently residing in the United Kingdom or the United States.

\subsubsection{Survey Structure}
To verify the robustness of our model in representing user concerns across various implementations of MR, four versions of the survey were created, with two versions introducing an AR prototype and two versions showing a VR prototype instead. Each version was shown to 25\% of the participant pool, ensuring equal distribution. Furthermore, one prototype per technology was described to feature functionality that is usually linked to be rather concerning, while the other prototype was selected to showcase features that are typically associated with lower levels of concern. This was done to ensure the scale could consistently gauge concerns across a spectrum of intensities for various types of MR technologies. They were each described with neutrally phrased text of roughly 200 to 300 words and a mockup image of the interface/system. The participants were asked to state how much they agreed with each item of the reduced item set on a 5-item Likert scale (Strongly disagree, Disagree, Neutral, Agree, Strongly agree).

All four prototypes were based on related work and already existing technologies. The non-concerning AR system was an intelligent navigation system, showing navigational clues via holograms and rerouting the user based on their preferences and current traffic information. This is based on already existing systems, implemented and tested in both research and industry environments~\cite{narztAugmentedRealityNavigation2006,bhorkarSurveyAugmentedReality2017}. The concerning AR system was based on "FlirtAR"\footnote{\url{https://flirtar.co}, last accessed on 2023-12-12.} and "ARR, matey!"\footnote{\url{https://wp.nyu.edu/tlt/conversational-roleplay-using-augmented-reality/}, last accessed on 2023-12-12.}, describing a dating app that would show information about the conversation partner and conversational suggestions via AR. The non-concerning VR system featured a virtual vacation application, similar to a multitude of readily available VR apps\footnote{\url{https://www.meta.com/de-de/blog/quest/virtual-vacation-11-vr-apps-and-films-that-let-you-travel-the-world-from-home/}, last accessed on 2023-12-12.} and related research~\cite{numazakiVREntertainmentSystem2017,mcleanDigitalTourismConsumption2023}. Lastly, the concerning VR prototype featured a gaming scenario, which would adapt the difficulty based on the player's emotions and physiological signals, porting the preexisting work of Chanel et al.~\cite{chanelEmotionAssessmentPhysiological2011} into a VR environment.

\subsection{Exploratory Factor Analysis}
Analogous to the development of other scales in the field of HCI~\cite{wozniakCreepyTechnologyWhat2021,mejiaNineItemQuestionnaireMeasuring2017,villaSocietyAttitudesHuman2022}, the extraction of latent factors was conducted as proposed by McCoach et al.~\cite{mccoachInstrumentDevelopmentAffective2013}. The results of the reversed items were inverted, and the Kaiser-Meyer-Olkin (KMO) criterion~\cite{shresthaFactorAnalysisTool2021} was evaluated. With KMO values above 0.8 indicating satisfactory sampling adequacy and the result being $\text{KMO} = 0.93$ for the present dataset, we continued with the factor analysis. For this, the parallel factors technique~\cite{hornRationaleTestNumber1965} was used in conjunction with a Scree plot~\cite{cattellScreeTestNumber1966} to find the optimal number of principal axis factors. A varimax rotation was applied, as this orthogonal rotation method produces independent factors, aiming to allow the later reduction of items that load on multiple factors at once~\cite{zhangFactorRotationStandard2015}.
Herein, the scree plot analysis indicated three factors to be the optimal solution for the items at hand.

To further reduce the set of items to achieve a concise scale that is practical for application in MR research, items with factor loadings below 0.40 were removed, as they are generally considered inadequate for such models~\cite{raykovIntroductionPsychometricTheory2011}. Items with significant cross-loadings were consequently removed as well. The final scale consists of 3 items per factor, leading to a number of 9 items in total. Cronbach's alphas, indicating the internal consistency of the (sub)scales, all show adequate consistency for the three factors, and the overall Cronbach's alpha of $\alpha = 0.85$ for the scale as a whole confirms that suitable items were retained~\cite{cortinaWhatCoefficientAlpha1993}. The Cronbach's alphas of the subscale and the factor loadings of the items are all shown in \Cref{table:final-quest}. The model displays a good fit with $\text{KMO} = 0.81$, a Tucker Lewis Index~\cite{caiIncrementalModelFit2023} of $\text{TLI} = 0.98$ and a Root Mean Square Error of Approximation of $\text{RMSEA} = 0.049$. 

\subsubsection{Factor Naming}
As the first three items (\textbf{SP1}, \textbf{SP2}, \textbf{SP3}) are a combination of the two subcategories \textit{Security} and \textit{Privacy}, we opted to name the first factor \textbf{Security \& Privacy}. This is in accordance with the work of De Guzman et al.~\cite{deguzmanSecurityPrivacyApproaches2019}, as introduced in \Cref{sec:secpriv}, where the two factors were also combined into one contiguous list of properties. The items \textbf{SI1}, \textbf{SI2}, and \textbf{SI3} all stem from the \textbf{Social Implications} subcategory of the conceptual framework, making the naming of the second factor trivial. Interestingly, the last three items (\textbf{T1}, \textbf{T2}, \textbf{T3}) all were reversed items. While their content in part correlates with the \textit{Security} and \textit{Privacy} categories, they also closely align with the last category, that being \textit{Public Acceptance}, or to be more precise, \textit{Trust}. As other properties of the \textit{Public Acceptance} are not present in the final item set anymore and with the first factor already covering the potential concerns regarding both \textit{Security} and \textit{Privacy}, we decided to name this factor \textbf{Trust}. With this factor only consisting of reversed items, we hope also to reduce the latent negative bias that might stem from the critically phrased items of the preceding factors. 

\begin{table*}[ht]
    \centering
    \caption{The final MRC Questionnaire, comprised of three factors with three items each. Also reported are Cronbach's alphas and factor loadings based on the first survey results.}
    \label{table:final-quest}
    \begin{tabular}{ M{0.65\textwidth} M{0.1\textwidth} M{0.03\textwidth} M{0.03\textwidth} M{0.03\textwidth}}
        \hline \multirow{2}{*}{\textbf{Subscale/Item}} & \multirow{2}{*}{\textbf{ID}} & \multicolumn{3}{c}{\textbf{Factor Loading}} \\\cline{3-5}
         & & \textbf{SP} & \textbf{SI} & \textbf{T} \\
        \hline \textbf{Security \& Privacy}, $\alpha = 0.88$ & & & & \\
        \hline 
        I am concerned about the possibility of non-authenticated individuals gaining access to this MR system. & \textbf{SP1} & 0.80 & & \\
        I am concerned about the potential exposure of sensitive data through this MR system to unauthorized parties. & \textbf{SP2} & 0.80 & & \\
        I worry that using this MR system might lead to my personal information being misused. & \textbf{SP3} & 0.80 & & \\
        \hline \textbf{Social Implications}, $\alpha = 0.81$ & & & & \\
        \hline 
        I fear that with this MR system, it becomes increasingly hard to maintain a clear distinction between virtual behavior and real-life behavior. & \textbf{SI1} & & 0.78 & \\
        I am concerned about the potential of this MR system to influence my behaviors in ways that could be detrimental to my well-being. & \textbf{SI2} & & 0.75 & \\
        Using this MR system might make me appear disconnected from others in my physical environment. & \textbf{SI3} & & 0.74 & \\
        \hline \textbf{Trust}, $\alpha = 0.88$ & & & & \\
        \hline 
        I believe that only legitimate individuals can access this MR system. (R) & \textbf{T1} & & & 0.77 \\
        I am sure that this MR system is maintaining a secure environment. (R) & \textbf{T2} & & & 0.78 \\
        I am confident that my anonymity is protected by this MR system. (R) & \textbf{T3} & & & 0.74 \\
        \hline 
    \end{tabular}
\end{table*}
\section{Scale Evaluation}
\label{sec:evaluation}
With the three final factors of the scale being determined, the MRC Questionnaire could now be evaluated appropriately. This process followed the \textit{Phase 3} of scale development by Boateng et al.~\cite{boatengBestPracticesDeveloping2018} and included two further surveys.

\subsection{Second Survey}
The first of the two surveys for evaluation was carried out to gather data for a confirmatory factor analysis, convergent/divergent validity, and differentiation by known groups.

\subsubsection{Participants}
As with the first survey, see \Cref{sec:1stsurvey}, we again chose to use Prolific as the recruitment platform for this survey. Similarly, the participation was entirely voluntary, and the option to withdraw from the survey was available throughout. Participants were compensated with \pounds1 upon completion of the survey, which corresponds to an average hourly reward of \pounds13.62. 
In total, $n = 100$ participants were recruited for this survey.
It was conducted entirely online and took approximately 5 minutes to complete. The average age of participants was again roughly 40 years ($\bar{x} = 39.83$, $s = 12.05$), 50\% identifying as male, 50\% identifying as female, and all either currently residing in the United Kingdom or the United States.

\subsubsection{Survey Structure}
Participants were shown one of two prototypes for assessment, one again being expected to yield comparatively few concerns and one raising potentially more concerns. Each was depicted using neutral-worded descriptions, spanning approximately 200 to 300 words, along with a mockup image of the interface/system. Each participant was randomly assigned one of the two prototypes, and both were shown with equal frequency.

Both systems offered the same fundamental feature set, namely an AR application offering contextual information for tourists in cities unfamiliar to them. This included the navigation to relevant points of interest through holograms that blend into the environment for unobtrusive clues, offering an adaptive AR experience. The second prototype introduced an additional feature, specifically blocking the view of parts of reality based on user preference. The hypothetical adaptive AR base system is based on related work~\cite{jingenliangSystematicReviewAugmented2021,damalaTailoringAdaptiveAugmented2012}, and the added view filter has been discussed in recent publications~\cite{eghtebasCoSpeculatingDarkScenarios2023} and expert interviews\footnote{\url{https://www.businessinsider.com/facebook-meta-metaverse-splinter-reality-more-2021-11}, last accessed on 2023-12-12.} as well.

After an introduction to the prototype at hand, participants were instructed to state their agreement with each of the items of the final MRC Questionnaire as shown in \Cref{table:final-quest}. Additionally, they were asked to complete both the PCTS~\cite{wozniakCreepyTechnologyWhat2021} and the UEQ~\cite{schreppConstructionBenchmarkUser2017a} for the shown prototype to facilitate convergent/divergent validity tests.

\subsection{Confirmatory Factor Analysis}
\label{sec:cfa}
\begin{figure*}[ht!]
    \centering
    \includegraphics[width=\textwidth]{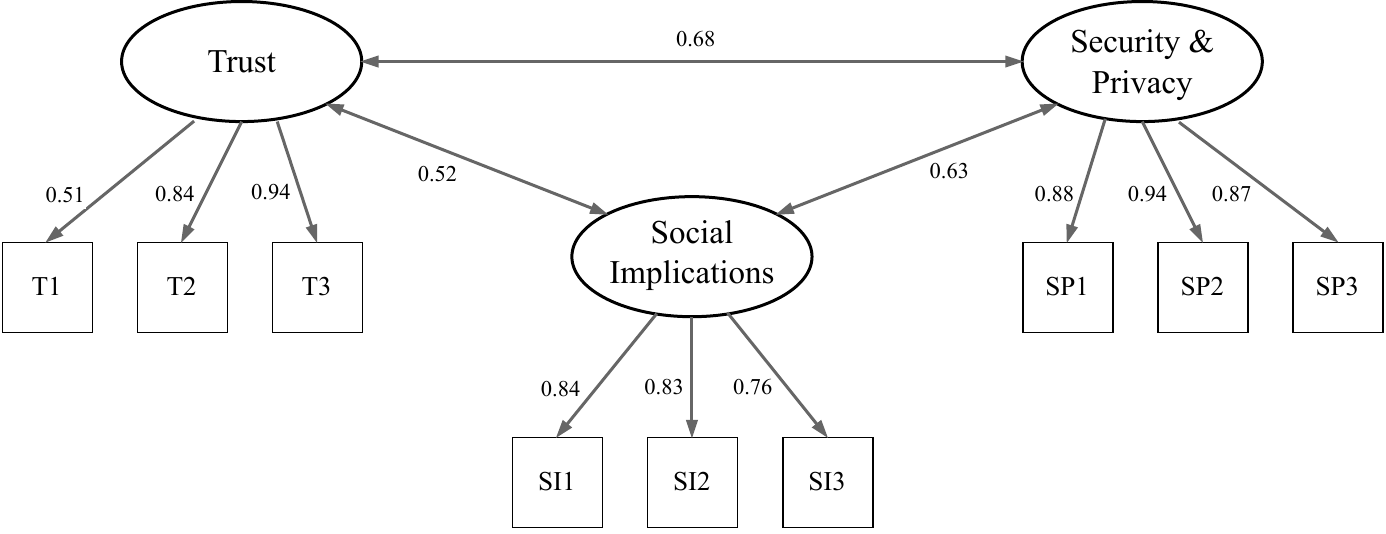}
    \caption{The result of the CFA confirms this three-factor model for the scale, with moderate correlations between the subscales and mostly high item coefficients.}
    \label{fig:lavaan}
    \Description{This figure shows the three-factor model as a graph. The three factors are split up into their sub-items. The sub-items are connected to their parent-factor and the correlations are given. Furthermore, the correlations between the three factors are given as well.}
\end{figure*}

To evaluate the structural validity of the scale, we performed a Confirmatory Factor Analysis (CFA). Herein, the dimensionality of the model can be verified through systematic fit assessments, confirming the structure of the model if certain thresholds are met~\cite{boatengBestPracticesDeveloping2018}. With a Tucker Lewis Index of $\text{TLI} = 0.98$, a Comparative Fit Index of $\text{CFI} = 0.99$, and a Root Mean Square Error of Approximation of $\text{RMSEA} = 0.059$ the results are indicative of an internally consistent model with a fair to close fit. As seen in \Cref{fig:lavaan}, the subscales exhibit a moderate to high correlation, implying that the theoretical scale is reasonable. The Cronbach's alphas for the three subscales are $\alpha = 0.92$ for \textit{Security \& Privacy}, $\alpha = 0.85$ for \textit{Social Implications}, and $\alpha = 0.79$ for \textit{Trust}, respectively. 

\subsection{Construct Validity}
As the two prototypes for the second survey were consciously chosen to differ in the number of concerns raised through having the same set of basic features, but the second one added a real-life filter that is already critically discussed in current literature, a differentiation by known groups is possible. Afterward, the MRC Questionnaire is compared to existing scales to evaluate if and how different concepts correlate with the proposed model.

\subsubsection{Differentiation by known groups}
The two prototypes for the second survey were intentionally selected to raise varying levels of concerns by sharing the same fundamental features, with the second introducing additional functionalities that have already been the subject of critical discussion in current literature. A differentiation by known groups can be performed on the assumption that the second prototype will cause significantly more concern among the participants. This approach was first proposed by Churchill et al.~\cite{churchillParadigmDevelopingBetter1979} and was analogously in previous scale development processes~\cite{wozniakCreepyTechnologyWhat2021,mejiaNineItemQuestionnaireMeasuring2017}.
The results of the second survey, divided into the two prototypes and analyzed separately, prove this assumption to be correct.
After assessing that a normal distribution could be assumed with a Shapiro-Wilk test ($\text{W} = 0.99$, $p = 0.34$) and that homogeneity of variances is given with Levene's test ($\text{L}(1,96) = 1.38, p = 0.24$), an independent t-test ($t(96) = -3.36, p = 0.001$) revealed that the resulting score of the MRC Questionnaire for the first scenario ($\bar{x}_\text{MRC} = 29.1, s_\text{MRC} = 6.96$) was significantly lower than for the second scenario ($\bar{x}_\text{MRC} = 33.6, s_\text{MRC} = 6.3$).
\Cref{table:diff-known} shows the full results of this step.


\begin{table*}[ht]
     \centering
     \caption{Differentiation by known groups.}
     \label{table:diff-known}
    \begin{tabular}{p{0.20\textwidth} M{0.08\textwidth} M{0.08\textwidth} M{0.08\textwidth} M{0.08\textwidth} M{0.25\textwidth}}
    \toprule
    \multirow{2}{*}{\textbf{Scale/Subscale}} & \multicolumn{2}{c}{\textbf{Scenario 1}} & \multicolumn{2}{c}{\textbf{Scenario 2}} & \multirow{2}{*}{\textbf{Independent t-Test}} \\
    \cmidrule(lr){2-3} \cmidrule(lr){4-5}
    & $\bar{x}$ & $s$ & $\bar{x}$ & $s$ & \\
    \midrule
    MRC                       & 29.1 & 6.96 & 33.6  & 6.3  & $t(96) = -3.36, p = 0.001$  \\
    Security \& Privacy       & 9.86 & 3.2 & 11.7  & 2.63  & $t(97) = -3.17, p = 0.002$  \\
    Social Implications       & 9.76 & 3.22 & 11.9  & 2.7 & $t(97) = -3.53, p < 0.001$  \\
    Trust                     & 9.49 & 2.19 & 10  & 2.45  & $t(97) = -1.09, p = 0.278$ \\
    \bottomrule
    \end{tabular}
\end{table*}

\subsubsection{Convergent/Divergent Validity}
\label{sec:condiv}
To compare the results from the MRC Questionnaire with established questionnaires, participants evaluated the presented hypothetical prototypes using not only the MRC but also the PCTS~\cite{wozniakCreepyTechnologyWhat2021} and the UEQ~\cite{schreppConstructionBenchmarkUser2017a}.

\begin{figure*}[ht!]
    \centering
    \includegraphics[width=\textwidth]{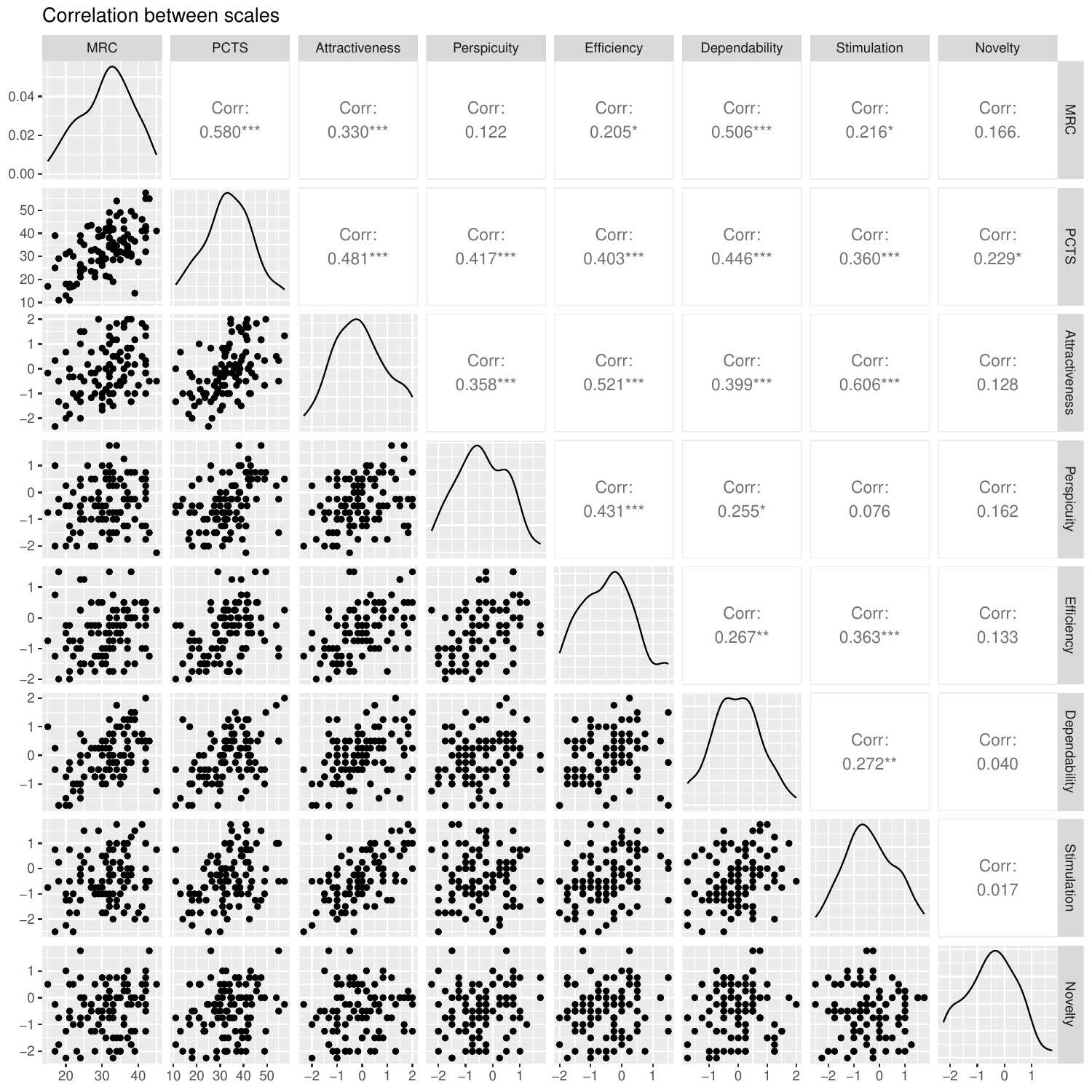}
    \caption{The main diagonal shows the histograms of each metric; the lower triangular shows the correlation plots between the metrics, and the upper triangular shows the corresponding r-values for the different scales under comparison. It is evident that the MRC and PCTS~\cite{wozniakCreepyTechnologyWhat2021} highly correlate. Furthermore, the MRC correlates with both the \textit{Attractiveness} and \textit{Dependability} subscales of the UEQ~\cite{schreppConstructionBenchmarkUser2017a}.}
    \label{fig:cor-mult}
    \Description{This figure shows an 8 by 8 matrix of smaller plots. The lower triangular shows 2D point clouds of the different correlations between the scales under test. The upper triangular shows just the calculated r-values for comparison. On the main diagonal, the histograms of each scale can be seen as line graphs.}
\end{figure*}

As the PCTS is one of the only scales that explicitly sets out to measure negative sentiments towards technologies, a high correlation between the MRC Questionnaire and it is desired.
The PCTS assesses the perceived creepiness of a technology in regards to the three factors \textit{Implied Malice}, \textit{Undesirability}, and \textit{Unpredicability}. One might assume that when individuals perceive a technology as having potential security or privacy vulnerabilities, they may consider it undesirable. The presence of security and privacy concerns might undermine the technology's trustworthiness, potentially making it less predictable in turn. Furthermore, when users perceive a technology as having social implications that may disrupt or harm societal norms, they may interpret these consequences as indicative of implied malice.
To the best of the authors' knowledge, there is currently no other questionnaire specifically designed to evaluate negative sentiments toward emerging technologies directly.
As \Cref{fig:cor-mult} shows, the MRC and PCTS correlate ($r = 0.58, \text{95\% CI} = 0.43, 0.70$), indicating that the perceived feeling of creepiness evoked by an MR system and the magnitude of concerns raised in relation to it are both impacted similarly. While a simple correlation test cannot prove the above-mentioned hypothesized causations, the scales do correlate as expected. While we assume that the PCTS assesses the feelings (i.e., invoked creepiness) that are a reaction to the system's concerns, and with this its inherent properties, further research is needed to prove this connection.

We incorporated the UEQ~\cite{schreppConstructionBenchmarkUser2017a} for another comparative assessment.
The comparison with the UEQ is particularly valuable due to its widespread use and established reputation as a comprehensive tool for assessing overall user experience, encompassing classical usability aspects as well as user experience dimensions. Among the available questionnaires, the UEQ was chosen for its versatility and applicability across various technological contexts, providing a well-established benchmark against which the effectiveness and specificity of the MRC questionnaire can be meaningfully evaluated.
While factors like efficiency or perspicuity can be hard to assess through a text description and a mockup image only, we specifically focused on the two hedonic qualities, those being stimulation and novelty.
Our interest in these hedonic qualities arises from the hypothesis that when a new device is perceived as subpar or unneeded, users may harbor more concerns. Conversely, when a new system is viewed as exceedingly novel and futuristic, concerns may stem more from unfamiliarity than from actual substantive concerns regarding the device. However, the test results reveal that both stimulation ($r = 0.22, \text{95\% CI} = 0.02, 0.39$) and novelty ($r = 0.17, \text{95\% CI} = -0.03, 0.35$) exhibit a low correlation with the MRC, suggesting that concerns related to MR systems encompass more than just stimulation and novelty. For completeness sake, all UEQ scales are shown in \Cref{fig:cor-mult}.


\subsection{Third Survey}
In addition to the Cronbach's alphas reported in \Cref{sec:cfa} as tests of reliability, we opted for performing one further test-retest reliability evaluation by conducting a final third survey.

\subsubsection{Participants}
Instead of using Prolific for recruitment, as in the first two surveys, participants were invited to take part through institute mailing lists and snowball sampling. In the end, a total of $n = 12$ people participated in the online survey, which took approximately 5 minutes to complete. Again, the participation was entirely voluntary, and the option to withdraw from the survey was available throughout. No compensation was given for the third survey. The average age of participants was roughly 27 years ($\bar{x} = 27.25$, $s = 4.0$), with two-thirds ($n = 8$) identifying as female and the rest identifying as male and all currently residing in countries of the European Union. As noted by Mejia and Yarosh~\cite{mejiaNineItemQuestionnaireMeasuring2017}, while it often poses difficulty to recruit enough people for two survey runs, and this usually being the reason why a test-retest evaluation is omitted, we too opted for still performing this validation, even if only a smaller sample size could be achieved. The time between the two runs was set to be at least ten days to ensure a long enough time between the two reflections on the presented prototype.

\subsubsection{Survey Structure}
Participants were shown one hypothetical prototype, for which, based on the explained feature set, relatively high values were to be expected. It consisted of an AR social application that enabled users to receive automatic information about their conversation partners through facial recognition. Additionally, it provided the functionality to rate individuals and conversations publicly. This concept was based on related work~\cite{isbisterHelperAgentDesigning2000,hoqueMyAutomatedConversation2012} and a now-defunct social media platform with a similar set of features\footnote{\url{https://techcrunch.com/2016/03/08/controversial-people-rating-app-peeple-goes-live-has-a-plan-to-profit-from-users-negative-reviews}, last accessed on 2023-12-12.}.

The MR system was described in a 260-word description, and a mockup of a potential interface for such an application was supplied. Afterward, participants were instructed to state their agreement with each of the items of the final MRC Questionnaire as shown in \Cref{table:final-quest}.

\subsection{Test-Retest Reliability}
\begin{figure*}[!ht]
    \begin{subfigure}{0.24\textwidth}
        \centering
        \includegraphics[width=\textwidth]{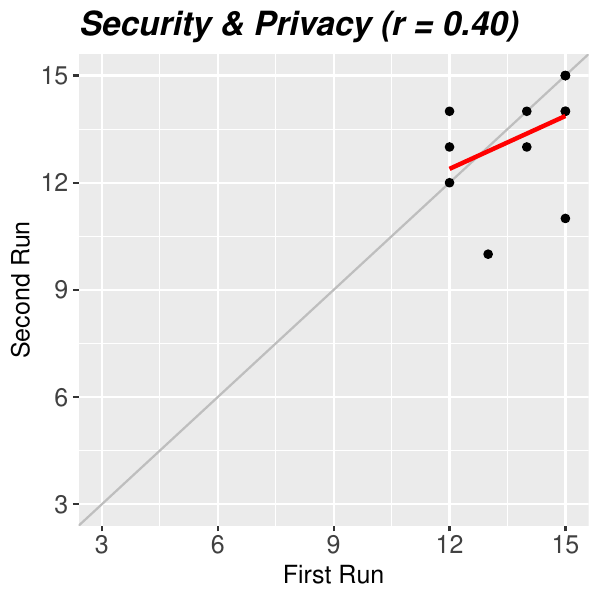}
    \end{subfigure}
    \begin{subfigure}{0.24\textwidth}
        \centering
        \includegraphics[width=\textwidth]{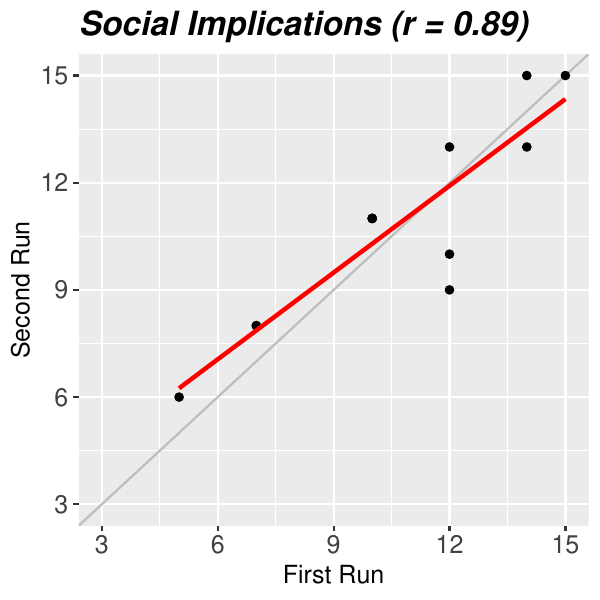}
    \end{subfigure}
    \begin{subfigure}{0.24\textwidth}
        \centering
        \includegraphics[width=\textwidth]{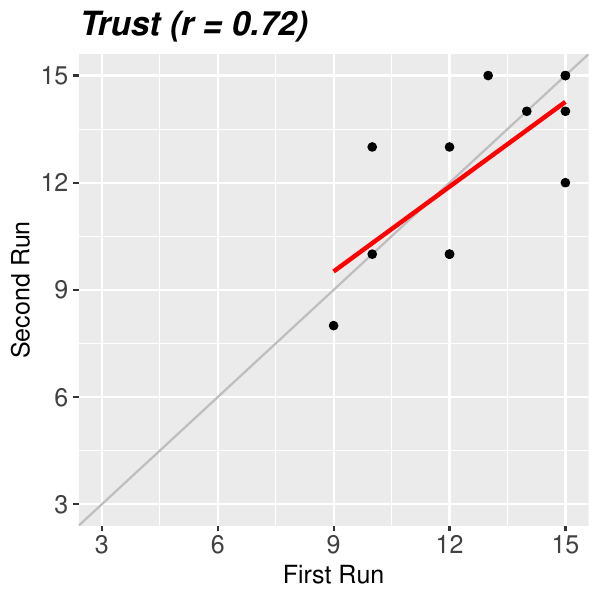}
    \end{subfigure}
    \begin{subfigure}{0.24\textwidth}
        \centering
        \includegraphics[width=\textwidth]{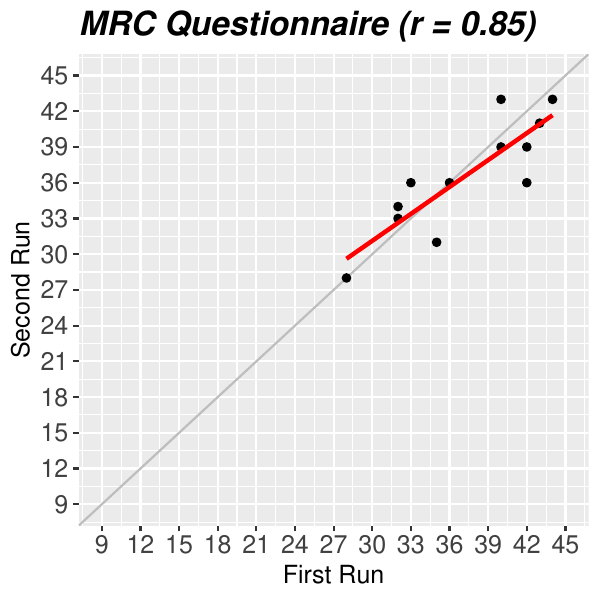}
    \end{subfigure}
    \caption{The different subscale and overall scores for both runs of the third survey. Furthermore, the Pearson product-moment correlation is given for all plots.}
     \label{fig:cor}
     \Description{In this figure, the test-retest results are shown for the subscales and the MRC questionnaire as a whole. Security and Privacy shows the lowest correlation between the two tests, all others show a relatively high correlation.}
\end{figure*}

As suggested by Rousson et al.~\cite{roussonAssessingIntraraterInterrater2002}, we evaluated the Pearson product-moment correlations for both the subscales and the MRC Questionnaire as a whole. While the \textit{Security \& Privacy} only showed an acceptable correlation for a test-retest context~\cite{cohenStatisticalPowerAnalysis2009}, the two other subscales showed much higher correlations. In total, the MRC Questionnaire exhibits a moderate to excellent test-retest reliability ($r = 0.85, \text{95\% CI} = 0.54, 0.96$). The correlation plots and respective correlation values are shown in \Cref{fig:cor}. Based on this reliability test, especially considering the small sample size, it can be assumed that the MRC Questionnaire shows temporal stability and can be used in repeated-measures studies.


\section{Discussion}
\label{sec:discussion}
In this section, we present instructions on using the MRC Questionnaire and interpreting its results. Furthermore, we explain the limitations of our approach and the scale as well as ideas for future enhancements.

\subsection{Scoring}
The MRC Questionnaire is scored on a 5-point Likert scale, ranging from \textit{Strongly disagree} (1) to \textit{Strongly agree} (5). All items of the \textit{Trust} subscale are reverse-coded.

\begin{align*}
    \text{MRC} &= \text{MRC}_\text{SP} + \text{MRC}_\text{SI} + \text{MRC}_\text{T}\\
    \text{with } \text{MRC}_\text{SP} &= \text{SP1} + \text{SP2} + \text{SP3}\\
    \text{and } \text{MRC}_\text{SI} &= \text{SI1} + \text{SI2} + \text{SI3}\\
    \text{and } \text{MRC}_\text{T} &= \text{T1}_R + \text{T2}_R + \text{T3}_R\\
\end{align*}

As a result, the scale's range spans from 9 as the lowest score to 45 as the highest. Elevated scores signify higher concerns associated with the MR system.

\subsection{Guidelines and Limitations to Administering the Scale}
A measuring instrument, such as the presented MRC Questionnaire, which is designed to assess concerns related to MR systems, can be immensely valuable for the research, development, and improvement of these technologies. Such an instrument might serve as a crucial tool in several ways:

This scale is intentionally designed not to assess the specific, objective problems or risks associated with a technology but rather to focus on user apprehensions and concerns. Its primary purpose is to measure the subjective perceptions and feelings of users regarding a technology, particularly any unease or worries they may experience.
By concentrating on user apprehensions, the scale aims to capture the emotional and psychological aspects of how MR systems might be perceived even before actual user experiences can be gathered. It recognizes that people's perceptions and concerns can vary widely, even when faced with similar objective risks or issues. Therefore, the scale provides a means to gauge how users interpret and respond to these risks on a personal level.

Conversely, it can also be used to assess actual implementations. Users' apprehensions often reveal pain points or areas of discomfort about the technology at hand. This information is valuable for pinpointing specific issues that may need addressing, whether they relate to security, privacy, social implications, or the inherent trust in the system. User concerns can also guide the development of educational materials or resources to help users understand the technology better. Addressing misconceptions or alleviating fears through education can contribute to a more positive user experience. In summary, while the scale's primary focus is on assessing user apprehensions and perceptions, it can serve as a versatile tool for evaluating new parts of the user experience in actual technology implementations, which other scales currently do not assess. By understanding and addressing user concerns, developers can enhance the overall quality and acceptance of MR systems and other technologies.

The preceding evaluation suggests that applying the MRC Questionnaire is suitable for both between-subject and within-subject studies, as well as for repeated-measures studies. Although the analysis of the subscales generally presents favorable results for evaluating them on their own, we do not explicitly recommend this application. The intentional brevity of the scale serves the purpose of offering a quick initial insight into potential user concerns. However, the precise nature of these concerns should be explored through additional qualitative research and is likely to be highly specific to the particular MR system under consideration. As illustrated by the conceptual model in \Cref{sec:concept}, the realm of potential reasons for concern is too expansive to encompass within a single scale suitable for a wide range of applications. Once again, this scale is designed primarily to provide an initial understanding of potential user concerns.

Finally, it is crucial to emphasize that this scale is not inherently linked to the acceptability of a system. Although we assume that the absence of concerns can certainly impact acceptability, numerous other factors may come into play. For this, other scales and questionnaires, like the ones presented in \Cref{sec:rel-quest} and \Cref{sec:condiv}, should be used in conjunction with the MRC Questionnaire.

\subsection{Limitations of the Development Process}
Next to the aforementioned limitations to how the scale can be used and evaluated, we acknowledge that the development process of the MRC Questionnaire may be subject to certain limitations, too. First and foremost, the exploratory factor analysis, as well as all subsequent evaluation stages, was conducted during a period when MR technologies were gradually making their way toward broader public acceptance. The trajectory of development and widespread adoption of these devices in the coming years remains uncertain. Consequently, it is likely that opinions, perceptions, and concerns will change over time. Therefore, a reevaluation of the scale may become necessary in the future.

Much like the PCTS~\cite{wozniakCreepyTechnologyWhat2021}, we opted to concentrate on developing a scale that evaluates users' concerns and apprehensions immediately after the first introduction to that MR system. Due to this, the suitability of the MRC for long-term studies remains uncertain. While we expect that the scale might have the potential to measure how user concerns change over time, it is essential to note that this capability cannot be definitively affirmed at the time being.

Additionally, the study primarily involved participants from countries with a Western cultural background, and as the surveys were conducted online, all participants possessed at least a basic understanding of current consumer electronics. While we hope for the scale to have relevance in diverse cultural contexts and among individuals with varying levels of familiarity with consumer electronics, we cannot guarantee this outcome. Ideally, future research will address this issue, facilitating cross-cultural and demographic comparisons of different concerns and apprehensions that people might have regarding MR systems.

The lack of real exposure testing introduces uncertainty of external factors (e.g., user context~\cite{4624429} or situatively perceived cognitive workload~\cite{koschSurveyMeasuringCognitive2023} during MR use), regarding the questionnaire's performance for capturing concerns when interacting with MR systems. The potential biases or deviations in user responses under actual MR exposure conditions raise consideration since they could impact the questionnaire's reliability and validity in such contexts (cf.~\cite{villa2023the, Boot2013Psychplac, kloft2023ai} for biased study data when users have specific expectations towards novel technologies). To address this limitation, future research should prioritize conducting evaluations with participants exposed to operational MR systems using the MRC questionnaire. This approach will provide a more comprehensive understanding of the MRC questionnaire's effectiveness in capturing user experiences. Additionally, incorporating user feedback from authentic MR interactions will contribute to refining the questionnaire for increased applicability and relevance in practical settings.
\section{Conclusion}
\label{sec:conclusion}
We present a measurement tool designed to evaluate user concerns and apprehensions regarding MR systems. Initially, we constructed a conceptual model outlining potential concerns associated with MR systems, drawing insights from existing research. Subsequently, we engaged in two rounds of expert feedback to generate a comprehensive set of survey items. 
A total of three surveys were conducted to first reduce this set of items and then evaluate the final MRC Questionnaire.

The questionnaire shows high internal consistency, adequate temporal stability, and high convergent and divergent validity. It serves as a valuable instrument for assessing the initial concerns individuals may harbor when encountering a new MR system. Furthermore, its intentional brevity enables its application in various studies and situations where an initial understanding of potential apprehensions is required.

We aspire for this scale to help researchers and developers cultivate a constructive approach to these concerns. It can serve as a tool to ensure that new MR artifacts and applications transparently convey their intentions, features, and potential impact on both users and bystanders. While this assessment could prove beneficial for educational purposes, it is essential to emphasize that addressing potential concerns primarily falls within the realm of technological development rather than solely relying on user adaptation or adjustment.

The questionnaire and supplementary material are openly accessible on the research group's website\footnote{\url{https://hcistudio.org/mrc-questionnaire}}.

\begin{acks}
This work was supported by the Swedish Research Council, award number 2022-03196.
\end{acks}


\bibliographystyle{ACM-Reference-Format}
\bibliography{main}


\begin{thebibliography}{79}


\ifx \showCODEN    \undefined \def \showCODEN     #1{\unskip}     \fi
\ifx \showDOI      \undefined \def \showDOI       #1{#1}\fi
\ifx \showISBNx    \undefined \def \showISBNx     #1{\unskip}     \fi
\ifx \showISBNxiii \undefined \def \showISBNxiii  #1{\unskip}     \fi
\ifx \showISSN     \undefined \def \showISSN      #1{\unskip}     \fi
\ifx \showLCCN     \undefined \def \showLCCN      #1{\unskip}     \fi
\ifx \shownote     \undefined \def \shownote      #1{#1}          \fi
\ifx \showarticletitle \undefined \def \showarticletitle #1{#1}   \fi
\ifx \showURL      \undefined \def \showURL       {\relax}        \fi
\providecommand\bibfield[2]{#2}
\providecommand\bibinfo[2]{#2}
\providecommand\natexlab[1]{#1}
\providecommand\showeprint[2][]{arXiv:#2}

\bibitem[Azmandian et~al\mbox{.}(2016)]%
        {azmandianHapticRetargetingDynamic2016}
\bibfield{author}{\bibinfo{person}{Mahdi Azmandian}, \bibinfo{person}{Mark Hancock}, \bibinfo{person}{Hrvoje Benko}, \bibinfo{person}{Eyal Ofek}, {and} \bibinfo{person}{Andrew~D. Wilson}.} \bibinfo{year}{2016}\natexlab{}.
\newblock \showarticletitle{Haptic {{Retargeting}}: {{Dynamic Repurposing}} of {{Passive Haptics}} for {{Enhanced Virtual Reality Experiences}}}. In \bibinfo{booktitle}{\emph{Proceedings of the 2016 {{CHI Conference}} on {{Human Factors}} in {{Computing Systems}}}} \emph{(\bibinfo{series}{{{CHI}} '16})}. \bibinfo{publisher}{{Association for Computing Machinery}}, \bibinfo{address}{{New York, NY, USA}}, \bibinfo{pages}{1968--1979}.
\newblock
\showISBNx{978-1-4503-3362-7}
\urldef\tempurl%
\url{https://doi.org/10.1145/2858036.2858226}
\showDOI{\tempurl}


\bibitem[Bhorkar(2017)]%
        {bhorkarSurveyAugmentedReality2017}
\bibfield{author}{\bibinfo{person}{Gaurav Bhorkar}.} \bibinfo{year}{2017}\natexlab{}.
\newblock \bibinfo{title}{A {{Survey}} of {{Augmented Reality Navigation}}}.
\newblock
\newblock
\urldef\tempurl%
\url{https://doi.org/10.48550/arXiv.1708.05006}
\showDOI{\tempurl}
\showeprint[arxiv]{1708.05006}~[cs]


\bibitem[Boateng et~al\mbox{.}(2018)]%
        {boatengBestPracticesDeveloping2018}
\bibfield{author}{\bibinfo{person}{Godfred~O. Boateng}, \bibinfo{person}{Torsten~B. Neilands}, \bibinfo{person}{Edward~A. Frongillo}, \bibinfo{person}{Hugo~R. {Melgar-Qui{\~n}onez}}, {and} \bibinfo{person}{Sera~L. Young}.} \bibinfo{year}{2018}\natexlab{}.
\newblock \showarticletitle{Best {{Practices}} for {{Developing}} and {{Validating Scales}} for {{Health}}, {{Social}}, and {{Behavioral Research}}: {{A Primer}}}.
\newblock \bibinfo{journal}{\emph{Frontiers in Public Health}}  \bibinfo{volume}{6} (\bibinfo{year}{2018}), \bibinfo{pages}{149}.
\newblock
\showISSN{2296-2565}


\bibitem[Boot et~al\mbox{.}(2013)]%
        {Boot2013Psychplac}
\bibfield{author}{\bibinfo{person}{Walter Boot}, \bibinfo{person}{Daniel Simons}, \bibinfo{person}{Cary Stothart}, {and} \bibinfo{person}{Cassie Stutts~Berry}.} \bibinfo{year}{2013}\natexlab{}.
\newblock \showarticletitle{The Pervasive Problem With Placebos in Psychology Why Active Control Groups Are Not Sufficient to Rule Out Placebo Effects}.
\newblock \bibinfo{journal}{\emph{Perspectives on Psychological Science}}  \bibinfo{volume}{8} (\bibinfo{date}{07} \bibinfo{year}{2013}), \bibinfo{pages}{445--454}.
\newblock
\urldef\tempurl%
\url{https://doi.org/10.1177/1745691613491271}
\showDOI{\tempurl}


\bibitem[Bunz et~al\mbox{.}(2021)]%
        {bunzTAMAVRTSDevelopment2021}
\bibfield{author}{\bibinfo{person}{Ulla Bunz}, \bibinfo{person}{Jonmichael Seibert}, {and} \bibinfo{person}{Joshua Hendrickse}.} \bibinfo{year}{2021}\natexlab{}.
\newblock \showarticletitle{From {{TAM}} to {{AVRTS}}: Development and Validation of the Attitudes toward {{Virtual Reality Technology Scale}}}.
\newblock \bibinfo{journal}{\emph{Virtual Reality}} \bibinfo{volume}{25}, \bibinfo{number}{1} (\bibinfo{date}{March} \bibinfo{year}{2021}), \bibinfo{pages}{31--41}.
\newblock
\showISSN{1434-9957}
\urldef\tempurl%
\url{https://doi.org/10.1007/s10055-020-00437-7}
\showDOI{\tempurl}


\bibitem[B\"{u}ttner et~al\mbox{.}(2017)]%
        {buettner2017design}
\bibfield{author}{\bibinfo{person}{Sebastian B\"{u}ttner}, \bibinfo{person}{Henrik Mucha}, \bibinfo{person}{Markus Funk}, \bibinfo{person}{Thomas Kosch}, \bibinfo{person}{Mario Aehnelt}, \bibinfo{person}{Sebastian Robert}, {and} \bibinfo{person}{Carsten R\"{o}cker}.} \bibinfo{year}{2017}\natexlab{}.
\newblock \showarticletitle{The Design Space of Augmented and Virtual Reality Applications for Assistive Environments in Manufacturing: A Visual Approach}. In \bibinfo{booktitle}{\emph{Proceedings of the 10th International Conference on PErvasive Technologies Related to Assistive Environments}} (Island of Rhodes, Greece) \emph{(\bibinfo{series}{PETRA '17})}. \bibinfo{publisher}{Association for Computing Machinery}, \bibinfo{address}{New York, NY, USA}, \bibinfo{pages}{433–440}.
\newblock
\showISBNx{9781450352277}
\urldef\tempurl%
\url{https://doi.org/10.1145/3056540.3076193}
\showDOI{\tempurl}


\bibitem[Cai et~al\mbox{.}(2023)]%
        {caiIncrementalModelFit2023}
\bibfield{author}{\bibinfo{person}{Li Cai}, \bibinfo{person}{Seung~Won Chung}, {and} \bibinfo{person}{Taehun Lee}.} \bibinfo{year}{2023}\natexlab{}.
\newblock \showarticletitle{Incremental {{Model Fit Assessment}} in the {{Case}} of {{Categorical Data}}: {{Tucker}}{\textendash}{{Lewis Index}} for {{Item Response Theory Modeling}}}.
\newblock \bibinfo{journal}{\emph{Prevention Science}} \bibinfo{volume}{24}, \bibinfo{number}{3} (\bibinfo{date}{April} \bibinfo{year}{2023}), \bibinfo{pages}{455--466}.
\newblock
\showISSN{1573-6695}
\urldef\tempurl%
\url{https://doi.org/10.1007/s11121-021-01253-4}
\showDOI{\tempurl}


\bibitem[Casey et~al\mbox{.}(2021)]%
        {caseyImmersiveVirtualReality2021}
\bibfield{author}{\bibinfo{person}{Peter Casey}, \bibinfo{person}{Ibrahim Baggili}, {and} \bibinfo{person}{Ananya Yarramreddy}.} \bibinfo{year}{2021}\natexlab{}.
\newblock \showarticletitle{Immersive {{Virtual Reality Attacks}} and the {{Human Joystick}}}.
\newblock \bibinfo{journal}{\emph{IEEE Transactions on Dependable and Secure Computing}} \bibinfo{volume}{18}, \bibinfo{number}{2} (\bibinfo{date}{March} \bibinfo{year}{2021}), \bibinfo{pages}{550--562}.
\newblock
\showISSN{1941-0018}
\urldef\tempurl%
\url{https://doi.org/10.1109/TDSC.2019.2907942}
\showDOI{\tempurl}


\bibitem[Cattell(1966)]%
        {cattellScreeTestNumber1966}
\bibfield{author}{\bibinfo{person}{Raymond~B. Cattell}.} \bibinfo{year}{1966}\natexlab{}.
\newblock \showarticletitle{The {{Scree Test For The Number Of Factors}}}.
\newblock \bibinfo{journal}{\emph{Multivariate Behavioral Research}} \bibinfo{volume}{1}, \bibinfo{number}{2} (\bibinfo{date}{April} \bibinfo{year}{1966}), \bibinfo{pages}{245--276}.
\newblock
\showISSN{0027-3171}
\urldef\tempurl%
\url{https://doi.org/10.1207/s15327906mbr0102_10}
\showDOI{\tempurl}


\bibitem[Chanel et~al\mbox{.}(2011)]%
        {chanelEmotionAssessmentPhysiological2011}
\bibfield{author}{\bibinfo{person}{G. Chanel}, \bibinfo{person}{C. Rebetez}, \bibinfo{person}{M. B{\'e}trancourt}, {and} \bibinfo{person}{T. Pun}.} \bibinfo{year}{2011}\natexlab{}.
\newblock \showarticletitle{Emotion {{Assessment From Physiological Signals}} for {{Adaptation}} of {{Game Difficulty}}}.
\newblock \bibinfo{journal}{\emph{IEEE Transactions on Systems, Man, and Cybernetics - Part A: Systems and Humans}} \bibinfo{volume}{41}, \bibinfo{number}{6} (\bibinfo{date}{Nov.} \bibinfo{year}{2011}), \bibinfo{pages}{1052--1063}.
\newblock
\showISSN{1083-4427, 1558-2426}
\urldef\tempurl%
\url{https://doi.org/10.1109/TSMCA.2011.2116000}
\showDOI{\tempurl}


\bibitem[Churchill(1979)]%
        {churchillParadigmDevelopingBetter1979}
\bibfield{author}{\bibinfo{person}{Gilbert~A. Churchill}.} \bibinfo{year}{1979}\natexlab{}.
\newblock \showarticletitle{A {{Paradigm}} for {{Developing Better Measures}} of {{Marketing Constructs}}}.
\newblock \bibinfo{journal}{\emph{Journal of Marketing Research}} \bibinfo{volume}{16}, \bibinfo{number}{1} (\bibinfo{year}{1979}), \bibinfo{pages}{64--73}.
\newblock
\showISSN{0022-2437}
\urldef\tempurl%
\url{https://doi.org/10.2307/3150876}
\showDOI{\tempurl}
\showeprint[jstor]{3150876}


\bibitem[Cohen(2009)]%
        {cohenStatisticalPowerAnalysis2009}
\bibfield{author}{\bibinfo{person}{Jacob Cohen}.} \bibinfo{year}{2009}\natexlab{}.
\newblock \bibinfo{booktitle}{\emph{Statistical Power Analysis for the Behavioral Sciences} (\bibinfo{edition}{2. ed., reprint} ed.)}.
\newblock \bibinfo{publisher}{{Psychology Press}}, \bibinfo{address}{{New York, NY}}.
\newblock
\showISBNx{978-0-8058-0283-2}


\bibitem[Comrey(1988)]%
        {comreyFactoranalyticMethodsScale1988}
\bibfield{author}{\bibinfo{person}{Andrew~L. Comrey}.} \bibinfo{year}{1988}\natexlab{}.
\newblock \showarticletitle{Factor-Analytic Methods of Scale Development in Personality and Clinical Psychology}.
\newblock \bibinfo{journal}{\emph{Journal of Consulting and Clinical Psychology}} \bibinfo{volume}{56}, \bibinfo{number}{5} (\bibinfo{year}{1988}), \bibinfo{pages}{754--761}.
\newblock
\showISSN{1939-2117}
\urldef\tempurl%
\url{https://doi.org/10.1037/0022-006X.56.5.754}
\showDOI{\tempurl}


\bibitem[Corbett et~al\mbox{.}(2024)]%
        {10339660}
\bibfield{author}{\bibinfo{person}{Matthew Corbett}, \bibinfo{person}{Brendan David-John}, \bibinfo{person}{Jiacheng Shang}, \bibinfo{person}{Y.~Charlie Hu}, {and} \bibinfo{person}{Bo Ji}.} \bibinfo{year}{2024}\natexlab{}.
\newblock \showarticletitle{Securing Bystander Privacy in Mixed Reality While Protecting the User Experience}.
\newblock \bibinfo{journal}{\emph{IEEE Security \& Privacy}} \bibinfo{volume}{22}, \bibinfo{number}{1} (\bibinfo{year}{2024}), \bibinfo{pages}{33--42}.
\newblock
\urldef\tempurl%
\url{https://doi.org/10.1109/MSEC.2023.3331649}
\showDOI{\tempurl}


\bibitem[Cortina(1993)]%
        {cortinaWhatCoefficientAlpha1993}
\bibfield{author}{\bibinfo{person}{Jose~M. Cortina}.} \bibinfo{year}{1993}\natexlab{}.
\newblock \showarticletitle{What Is Coefficient Alpha? {{An}} Examination of Theory and Applications}.
\newblock \bibinfo{journal}{\emph{Journal of Applied Psychology}} \bibinfo{volume}{78}, \bibinfo{number}{1} (\bibinfo{year}{1993}), \bibinfo{pages}{98--104}.
\newblock
\showISSN{1939-1854}
\urldef\tempurl%
\url{https://doi.org/10.1037/0021-9010.78.1.98}
\showDOI{\tempurl}


\bibitem[Damala and Stojanovic(2012)]%
        {damalaTailoringAdaptiveAugmented2012}
\bibfield{author}{\bibinfo{person}{Areti Damala} {and} \bibinfo{person}{Nenad Stojanovic}.} \bibinfo{year}{2012}\natexlab{}.
\newblock \showarticletitle{Tailoring the {{Adaptive Augmented Reality}} ({{A2R}}) {{Museum Visit}}: {{Identifying Cultural Heritage Professionals}}' {{Motivations}} and {{Needs}}}. In \bibinfo{booktitle}{\emph{Nternational {{Symposium}} on {{Mixed}} and {{Augmented Reality}} 2012 ({{ISMAR}} 2012)}}. \bibinfo{publisher}{{ISMAR-AMH}}, \bibinfo{address}{{Atlanta, USA}}, \bibinfo{pages}{71--80}.
\newblock


\bibitem[Davis(1989)]%
        {davisPerceivedUsefulnessPerceived1989}
\bibfield{author}{\bibinfo{person}{Fred~D. Davis}.} \bibinfo{year}{1989}\natexlab{}.
\newblock \showarticletitle{Perceived {{Usefulness}}, {{Perceived Ease}} of {{Use}}, and {{User Acceptance}} of {{Information Technology}}}.
\newblock \bibinfo{journal}{\emph{MIS Quarterly}} \bibinfo{volume}{13}, \bibinfo{number}{3} (\bibinfo{year}{1989}), \bibinfo{pages}{319--340}.
\newblock
\showISSN{0276-7783}
\urldef\tempurl%
\url{https://doi.org/10.2307/249008}
\showDOI{\tempurl}
\showeprint[jstor]{249008}


\bibitem[Davis et~al\mbox{.}(1989)]%
        {davisUserAcceptanceComputer1989}
\bibfield{author}{\bibinfo{person}{Fred~D. Davis}, \bibinfo{person}{Richard~P. Bagozzi}, {and} \bibinfo{person}{Paul~R. Warshaw}.} \bibinfo{year}{1989}\natexlab{}.
\newblock \showarticletitle{User {{Acceptance}} of {{Computer Technology}}: {{A Comparison}} of {{Two Theoretical Models}}}.
\newblock \bibinfo{journal}{\emph{Management Science}} \bibinfo{volume}{35}, \bibinfo{number}{8} (\bibinfo{year}{1989}), \bibinfo{pages}{982--1003}.
\newblock
\showISSN{0025-1909}
\showeprint[jstor]{2632151}


\bibitem[De~Guzman et~al\mbox{.}(2019)]%
        {deguzmanSecurityPrivacyApproaches2019}
\bibfield{author}{\bibinfo{person}{Jaybie~A. De~Guzman}, \bibinfo{person}{Kanchana Thilakarathna}, {and} \bibinfo{person}{Aruna Seneviratne}.} \bibinfo{year}{2019}\natexlab{}.
\newblock \showarticletitle{Security and {{Privacy Approaches}} in {{Mixed Reality}}: {{A Literature Survey}}}.
\newblock \bibinfo{journal}{\emph{Comput. Surveys}} \bibinfo{volume}{52}, \bibinfo{number}{6} (\bibinfo{date}{Oct.} \bibinfo{year}{2019}), \bibinfo{pages}{110:1--110:37}.
\newblock
\showISSN{0360-0300}
\urldef\tempurl%
\url{https://doi.org/10.1145/3359626}
\showDOI{\tempurl}


\bibitem[Deng et~al\mbox{.}(2011)]%
        {dengPrivacyThreatAnalysis2011}
\bibfield{author}{\bibinfo{person}{Mina Deng}, \bibinfo{person}{Kim Wuyts}, \bibinfo{person}{Riccardo Scandariato}, \bibinfo{person}{Bart Preneel}, {and} \bibinfo{person}{Wouter Joosen}.} \bibinfo{year}{2011}\natexlab{}.
\newblock \showarticletitle{A Privacy Threat Analysis Framework: Supporting the Elicitation and Fulfillment of Privacy Requirements}.
\newblock \bibinfo{journal}{\emph{Requirements Engineering}} \bibinfo{volume}{16}, \bibinfo{number}{1} (\bibinfo{date}{March} \bibinfo{year}{2011}), \bibinfo{pages}{3--32}.
\newblock
\showISSN{1432-010X}
\urldef\tempurl%
\url{https://doi.org/10.1007/s00766-010-0115-7}
\showDOI{\tempurl}


\bibitem[Eghtebas et~al\mbox{.}(2023)]%
        {eghtebasCoSpeculatingDarkScenarios2023}
\bibfield{author}{\bibinfo{person}{Chloe Eghtebas}, \bibinfo{person}{Gudrun Klinker}, \bibinfo{person}{Susanne Boll}, {and} \bibinfo{person}{Marion Koelle}.} \bibinfo{year}{2023}\natexlab{}.
\newblock \showarticletitle{Co-{{Speculating}} on {{Dark Scenarios}} and {{Unintended Consequences}} of a {{Ubiquitous}}(Ly) {{Augmented Reality}}}. In \bibinfo{booktitle}{\emph{Proceedings of the 2023 {{ACM Designing Interactive Systems Conference}}}} \emph{(\bibinfo{series}{{{DIS}} '23})}. \bibinfo{publisher}{{Association for Computing Machinery}}, \bibinfo{address}{{New York, NY, USA}}, \bibinfo{pages}{2392--2407}.
\newblock
\showISBNx{978-1-4503-9893-0}
\urldef\tempurl%
\url{https://doi.org/10.1145/3563657.3596073}
\showDOI{\tempurl}


\bibitem[Evangelista~Belo et~al\mbox{.}(2021)]%
        {10.1145/3411764.3445349}
\bibfield{author}{\bibinfo{person}{Jo\~{a}o~Marcelo Evangelista~Belo}, \bibinfo{person}{Anna~Maria Feit}, \bibinfo{person}{Tiare Feuchtner}, {and} \bibinfo{person}{Kaj Gr\o{}nb\ae{}k}.} \bibinfo{year}{2021}\natexlab{}.
\newblock \showarticletitle{XRgonomics: Facilitating the Creation of Ergonomic 3D Interfaces}. In \bibinfo{booktitle}{\emph{Proceedings of the 2021 CHI Conference on Human Factors in Computing Systems}} \emph{(\bibinfo{series}{CHI '21})}. \bibinfo{publisher}{Association for Computing Machinery}, \bibinfo{address}{New York, NY, USA}, Article \bibinfo{articleno}{290}, \bibinfo{numpages}{11}~pages.
\newblock
\showISBNx{9781450380966}
\urldef\tempurl%
\url{https://doi.org/10.1145/3411764.3445349}
\showDOI{\tempurl}


\bibitem[Feger et~al\mbox{.}(2022)]%
        {feger2022electronicsar}
\bibfield{author}{\bibinfo{person}{Sebastian Feger}, \bibinfo{person}{Lars Semmler}, \bibinfo{person}{Albrecht Schmidt}, {and} \bibinfo{person}{Thomas Kosch}.} \bibinfo{year}{2022}\natexlab{}.
\newblock \showarticletitle{{ElectronicsAR: Design and Evaluation of a Mobile and Tangible High-Fidelity Augmented Electronics Toolkit}}.
\newblock \bibinfo{journal}{\emph{{Proc. ACM Hum.-Comput. Interact.}}} \bibinfo{volume}{6}, \bibinfo{number}{ISS}, Article \bibinfo{articleno}{587} (\bibinfo{date}{nov} \bibinfo{year}{2022}), \bibinfo{numpages}{22}~pages.
\newblock
\urldef\tempurl%
\url{https://doi.org/10.1145/3567740}
\showDOI{\tempurl}


\bibitem[George et~al\mbox{.}(2008)]%
        {georgeMeasuringImplementationSchools2008}
\bibfield{author}{\bibinfo{person}{Archie~A. George}, \bibinfo{person}{Gene~E. Hall}, {and} \bibinfo{person}{Suzanne Stiegelbauer}.} \bibinfo{year}{2008}\natexlab{}.
\newblock \bibinfo{booktitle}{\emph{Measuring Implementation in Schools: The Stages of Concern Questionnaire} (\bibinfo{edition}{2. print. with minor additions and corr} ed.)}.
\newblock \bibinfo{publisher}{{Southwest Educational Development Laboratory}}, \bibinfo{address}{{Austin, Tex}}.
\newblock
\showISBNx{978-0-9777208-0-4}


\bibitem[Georgiou and Kyza(2017)]%
        {georgiouDevelopmentValidationARI2017}
\bibfield{author}{\bibinfo{person}{Yiannis Georgiou} {and} \bibinfo{person}{Eleni~A. Kyza}.} \bibinfo{year}{2017}\natexlab{}.
\newblock \showarticletitle{The Development and Validation of the {{ARI}} Questionnaire}.
\newblock \bibinfo{journal}{\emph{International Journal of Human-Computer Studies}} \bibinfo{volume}{98}, \bibinfo{number}{C} (\bibinfo{date}{Feb.} \bibinfo{year}{2017}), \bibinfo{pages}{24--37}.
\newblock
\showISSN{1071-5819}
\urldef\tempurl%
\url{https://doi.org/10.1016/j.ijhcs.2016.09.014}
\showDOI{\tempurl}


\bibitem[Grier et~al\mbox{.}(2013)]%
        {grierSystemUsabilityScale2013}
\bibfield{author}{\bibinfo{person}{Rebecca~A. Grier}, \bibinfo{person}{Aaron Bangor}, \bibinfo{person}{Philip Kortum}, {and} \bibinfo{person}{S.~Camille Peres}.} \bibinfo{year}{2013}\natexlab{}.
\newblock \showarticletitle{The {{System Usability Scale}}: {{Beyond Standard Usability Testing}}}.
\newblock \bibinfo{journal}{\emph{Proceedings of the Human Factors and Ergonomics Society Annual Meeting}} \bibinfo{volume}{57}, \bibinfo{number}{1} (\bibinfo{date}{Sept.} \bibinfo{year}{2013}), \bibinfo{pages}{187--191}.
\newblock
\showISSN{2169-5067}
\urldef\tempurl%
\url{https://doi.org/10.1177/1541931213571042}
\showDOI{\tempurl}


\bibitem[Gugenheimer et~al\mbox{.}(2022)]%
        {gugenheimerNovelChallengesSafety2022}
\bibfield{author}{\bibinfo{person}{Jan Gugenheimer}, \bibinfo{person}{Wen-Jie Tseng}, \bibinfo{person}{Abraham~Hani Mhaidli}, \bibinfo{person}{Jan~Ole Rixen}, \bibinfo{person}{Mark McGill}, \bibinfo{person}{Michael Nebeling}, \bibinfo{person}{Mohamed Khamis}, \bibinfo{person}{Florian Schaub}, {and} \bibinfo{person}{Sanchari Das}.} \bibinfo{year}{2022}\natexlab{}.
\newblock \showarticletitle{Novel {{Challenges}} of {{Safety}}, {{Security}} and {{Privacy}} in {{Extended Reality}}}. In \bibinfo{booktitle}{\emph{Extended {{Abstracts}} of the 2022 {{CHI Conference}} on {{Human Factors}} in {{Computing Systems}}}} \emph{(\bibinfo{series}{{{CHI EA}} '22})}. \bibinfo{publisher}{{Association for Computing Machinery}}, \bibinfo{address}{{New York, NY, USA}}, \bibinfo{pages}{1--5}.
\newblock
\showISBNx{978-1-4503-9156-6}
\urldef\tempurl%
\url{https://doi.org/10.1145/3491101.3503741}
\showDOI{\tempurl}


\bibitem[Guo et~al\mbox{.}(2021)]%
        {guoSafetyHealthPerceptions2021}
\bibfield{author}{\bibinfo{person}{Yuntao Guo}, \bibinfo{person}{Shubham Agrawal}, \bibinfo{person}{Srinivas Peeta}, {and} \bibinfo{person}{Irina Benedyk}.} \bibinfo{year}{2021}\natexlab{}.
\newblock \showarticletitle{Safety and Health Perceptions of Location-Based Augmented Reality Gaming App and Their Implications}.
\newblock \bibinfo{journal}{\emph{Accident Analysis \& Prevention}}  \bibinfo{volume}{161} (\bibinfo{date}{Oct.} \bibinfo{year}{2021}), \bibinfo{pages}{106354}.
\newblock
\showISSN{0001-4575}
\urldef\tempurl%
\url{https://doi.org/10.1016/j.aap.2021.106354}
\showDOI{\tempurl}


\bibitem[Gupta et~al\mbox{.}(2012)]%
        {guptaSociopsychologicalDeterminantsPublic2012}
\bibfield{author}{\bibinfo{person}{Nidhi Gupta}, \bibinfo{person}{Arnout~R.H. Fischer}, {and} \bibinfo{person}{Lynn~J. Frewer}.} \bibinfo{year}{2012}\natexlab{}.
\newblock \showarticletitle{Socio-Psychological Determinants of Public Acceptance of Technologies: {{A}} Review}.
\newblock \bibinfo{journal}{\emph{Public Understanding of Science}} \bibinfo{volume}{21}, \bibinfo{number}{7} (\bibinfo{date}{Oct.} \bibinfo{year}{2012}), \bibinfo{pages}{782--795}.
\newblock
\showISSN{0963-6625}
\urldef\tempurl%
\url{https://doi.org/10.1177/0963662510392485}
\showDOI{\tempurl}


\bibitem[Harborth and Pape(2021)]%
        {harborthInvestigatingPrivacyConcerns2021}
\bibfield{author}{\bibinfo{person}{David Harborth} {and} \bibinfo{person}{Sebastian Pape}.} \bibinfo{year}{2021}\natexlab{}.
\newblock \showarticletitle{Investigating Privacy Concerns Related to Mobile Augmented Reality {{Apps}} {\textendash} {{A}} Vignette Based Online Experiment}.
\newblock \bibinfo{journal}{\emph{Computers in Human Behavior}}  \bibinfo{volume}{122} (\bibinfo{date}{Sept.} \bibinfo{year}{2021}), \bibinfo{pages}{106833}.
\newblock
\showISSN{0747-5632}
\urldef\tempurl%
\url{https://doi.org/10.1016/j.chb.2021.106833}
\showDOI{\tempurl}


\bibitem[Hassenzahl et~al\mbox{.}(2003)]%
        {hassenzahlAttrakDiffFragebogenZur2003}
\bibfield{author}{\bibinfo{person}{Marc Hassenzahl}, \bibinfo{person}{Michael Burmester}, {and} \bibinfo{person}{Franz Koller}.} \bibinfo{year}{2003}\natexlab{}.
\newblock \showarticletitle{{AttrakDiff: Ein Fragebogen zur Messung wahrgenommener hedonischer und pragmatischer Qualit{\"a}t}}.
\newblock In \bibinfo{booktitle}{\emph{{Mensch \& Computer 2003: Interaktion in Bewegung}}}, \bibfield{editor}{\bibinfo{person}{Gerd Szwillus} {and} \bibinfo{person}{J{\"u}rgen Ziegler}} (Eds.). \bibinfo{publisher}{{Vieweg+Teubner Verlag}}, \bibinfo{address}{{Wiesbaden}}, \bibinfo{pages}{187--196}.
\newblock
\showISBNx{978-3-322-80058-9}
\urldef\tempurl%
\url{https://doi.org/10.1007/978-3-322-80058-9_19}
\showDOI{\tempurl}


\bibitem[Hoque(2012)]%
        {hoqueMyAutomatedConversation2012}
\bibfield{author}{\bibinfo{person}{Mohammed~E. Hoque}.} \bibinfo{year}{2012}\natexlab{}.
\newblock \showarticletitle{My Automated Conversation Helper ({{MACH}}): Helping People Improve Social Skills.}. In \bibinfo{booktitle}{\emph{Proceedings of the 14th {{ACM}} International Conference on {{Multimodal}} Interaction}}. \bibinfo{publisher}{{ACM}}, \bibinfo{address}{{Santa Monica California USA}}, \bibinfo{pages}{313--316}.
\newblock
\showISBNx{978-1-4503-1467-1}
\urldef\tempurl%
\url{https://doi.org/10.1145/2388676.2388745}
\showDOI{\tempurl}


\bibitem[Horn(1965)]%
        {hornRationaleTestNumber1965}
\bibfield{author}{\bibinfo{person}{John~L. Horn}.} \bibinfo{year}{1965}\natexlab{}.
\newblock \showarticletitle{A Rationale and Test for the Number of Factors in Factor Analysis}.
\newblock \bibinfo{journal}{\emph{Psychometrika}} \bibinfo{volume}{30}, \bibinfo{number}{2} (\bibinfo{date}{June} \bibinfo{year}{1965}), \bibinfo{pages}{179--185}.
\newblock
\showISSN{1860-0980}
\urldef\tempurl%
\url{https://doi.org/10.1007/BF02289447}
\showDOI{\tempurl}


\bibitem[Howard and Lipner(2006)]%
        {howardSecurityDevelopmentLifecycle2006}
\bibfield{author}{\bibinfo{person}{Michael Howard} {and} \bibinfo{person}{Steve Lipner}.} \bibinfo{year}{2006}\natexlab{}.
\newblock \bibinfo{booktitle}{\emph{The {{Security Development Lifecycle}}: {{SDL}}, a {{Process}} for {{Developing Demonstrably More Secure Software}}}}.
\newblock \bibinfo{publisher}{{Microsoft Press}}, \bibinfo{address}{{Redmond, WA}}.
\newblock
\showISBNx{978-0-7356-2214-2}


\bibitem[Hughes et~al\mbox{.}(2005)]%
        {hughesMixedRealityEducation2005}
\bibfield{author}{\bibinfo{person}{C.E. Hughes}, \bibinfo{person}{C.B. Stapleton}, \bibinfo{person}{D.E. Hughes}, {and} \bibinfo{person}{E.M. Smith}.} \bibinfo{year}{2005}\natexlab{}.
\newblock \showarticletitle{Mixed Reality in Education, Entertainment, and Training}.
\newblock \bibinfo{journal}{\emph{IEEE Computer Graphics and Applications}} \bibinfo{volume}{25}, \bibinfo{number}{6} (\bibinfo{date}{Nov.} \bibinfo{year}{2005}), \bibinfo{pages}{24--30}.
\newblock
\showISSN{1558-1756}
\urldef\tempurl%
\url{https://doi.org/10.1109/MCG.2005.139}
\showDOI{\tempurl}


\bibitem[Hussain et~al\mbox{.}(2023)]%
        {hussainAugmentedRealitySickness2023}
\bibfield{author}{\bibinfo{person}{Muhammad Hussain}, \bibinfo{person}{Jaehyun Park}, {and} \bibinfo{person}{Hyun~K. Kim}.} \bibinfo{year}{2023}\natexlab{}.
\newblock \showarticletitle{Augmented Reality Sickness Questionnaire ({{ARSQ}}): {{A}} Refined Questionnaire for Augmented Reality Environment}.
\newblock \bibinfo{journal}{\emph{International Journal of Industrial Ergonomics}}  \bibinfo{volume}{97} (\bibinfo{date}{Sept.} \bibinfo{year}{2023}), \bibinfo{pages}{103495}.
\newblock
\showISSN{0169-8141}
\urldef\tempurl%
\url{https://doi.org/10.1016/j.ergon.2023.103495}
\showDOI{\tempurl}


\bibitem[Isbister et~al\mbox{.}(2000)]%
        {isbisterHelperAgentDesigning2000}
\bibfield{author}{\bibinfo{person}{Katherine Isbister}, \bibinfo{person}{Hideyuki Nakanishi}, \bibinfo{person}{Toru Ishida}, {and} \bibinfo{person}{Cliff Nass}.} \bibinfo{year}{2000}\natexlab{}.
\newblock \showarticletitle{Helper Agent: Designing an Assistant for Human-Human Interaction in a Virtual Meeting Space}. In \bibinfo{booktitle}{\emph{Proceedings of the {{SIGCHI}} Conference on {{Human Factors}} in {{Computing Systems}}}} \emph{(\bibinfo{series}{{{CHI}} '00})}. \bibinfo{publisher}{{Association for Computing Machinery}}, \bibinfo{address}{{New York, NY, USA}}, \bibinfo{pages}{57--64}.
\newblock
\showISBNx{978-1-58113-216-8}
\urldef\tempurl%
\url{https://doi.org/10.1145/332040.332407}
\showDOI{\tempurl}


\bibitem[Jian et~al\mbox{.}(2000)]%
        {jianFoundationsEmpiricallyDetermined2000}
\bibfield{author}{\bibinfo{person}{Jiun-Yin Jian}, \bibinfo{person}{Ann Bisantz}, {and} \bibinfo{person}{Colin Drury}.} \bibinfo{year}{2000}\natexlab{}.
\newblock \showarticletitle{Foundations for an {{Empirically Determined Scale}} of {{Trust}} in {{Automated Systems}}}.
\newblock \bibinfo{journal}{\emph{International Journal of Cognitive Ergonomics}}  \bibinfo{volume}{4} (\bibinfo{date}{March} \bibinfo{year}{2000}), \bibinfo{pages}{53--71}.
\newblock
\urldef\tempurl%
\url{https://doi.org/10.1207/S15327566IJCE0401_04}
\showDOI{\tempurl}


\bibitem[Jingen~Liang and Elliot(2021)]%
        {jingenliangSystematicReviewAugmented2021}
\bibfield{author}{\bibinfo{person}{Lena Jingen~Liang} {and} \bibinfo{person}{Statia Elliot}.} \bibinfo{year}{2021}\natexlab{}.
\newblock \showarticletitle{A Systematic Review of Augmented Reality Tourism Research: {{What}} Is Now and What Is Next?}
\newblock \bibinfo{journal}{\emph{Tourism and Hospitality Research}} \bibinfo{volume}{21}, \bibinfo{number}{1} (\bibinfo{date}{Jan.} \bibinfo{year}{2021}), \bibinfo{pages}{15--30}.
\newblock
\showISSN{1467-3584}
\urldef\tempurl%
\url{https://doi.org/10.1177/1467358420941913}
\showDOI{\tempurl}


\bibitem[Kalloniatis et~al\mbox{.}(2008)]%
        {kalloniatisAddressingPrivacyRequirements2008}
\bibfield{author}{\bibinfo{person}{Christos Kalloniatis}, \bibinfo{person}{Evangelia Kavakli}, {and} \bibinfo{person}{Stefanos Gritzalis}.} \bibinfo{year}{2008}\natexlab{}.
\newblock \showarticletitle{Addressing Privacy Requirements in System Design: The {{PriS}} Method}.
\newblock \bibinfo{journal}{\emph{Requirements Engineering}} \bibinfo{volume}{13}, \bibinfo{number}{3} (\bibinfo{date}{Sept.} \bibinfo{year}{2008}), \bibinfo{pages}{241--255}.
\newblock
\showISSN{1432-010X}
\urldef\tempurl%
\url{https://doi.org/10.1007/s00766-008-0067-3}
\showDOI{\tempurl}


\bibitem[Katins et~al\mbox{.}(2023a)]%
        {10.1145/3544549.3585742}
\bibfield{author}{\bibinfo{person}{Christopher Katins}, \bibinfo{person}{Sebastian~S. Feger}, {and} \bibinfo{person}{Thomas Kosch}.} \bibinfo{year}{2023}\natexlab{a}.
\newblock \showarticletitle{Exploring Mixed Reality in General Aviation to Support Pilot Workload}. In \bibinfo{booktitle}{\emph{Extended Abstracts of the 2023 CHI Conference on Human Factors in Computing Systems}} (Hamburg, Germany) \emph{(\bibinfo{series}{CHI EA '23})}. \bibinfo{publisher}{Association for Computing Machinery}, \bibinfo{address}{New York, NY, USA}, Article \bibinfo{articleno}{116}, \bibinfo{numpages}{7}~pages.
\newblock
\showISBNx{9781450394222}
\urldef\tempurl%
\url{https://doi.org/10.1145/3544549.3585742}
\showDOI{\tempurl}


\bibitem[Katins et~al\mbox{.}(2023b)]%
        {10.1145/3626705.3627785}
\bibfield{author}{\bibinfo{person}{Christopher Katins}, \bibinfo{person}{Sebastian~Stefan Feger}, {and} \bibinfo{person}{Thomas Kosch}.} \bibinfo{year}{2023}\natexlab{b}.
\newblock \showarticletitle{Pilots' Considerations Regarding Current Generation Mixed Reality Headset Use in General Aviation Cockpits}. In \bibinfo{booktitle}{\emph{Proceedings of the 22nd International Conference on Mobile and Ubiquitous Multimedia}} (Vienna, Austria) \emph{(\bibinfo{series}{MUM '23})}. \bibinfo{publisher}{Association for Computing Machinery}, \bibinfo{address}{New York, NY, USA}, \bibinfo{pages}{159–165}.
\newblock
\showISBNx{9798400709210}
\urldef\tempurl%
\url{https://doi.org/10.1145/3626705.3627785}
\showDOI{\tempurl}


\bibitem[Kaufeld et~al\mbox{.}(2022)]%
        {kaufeldOpticalSeethroughAugmented2022}
\bibfield{author}{\bibinfo{person}{Mara Kaufeld}, \bibinfo{person}{Martin Mundt}, \bibinfo{person}{Sarah Forst}, {and} \bibinfo{person}{Heiko Hecht}.} \bibinfo{year}{2022}\natexlab{}.
\newblock \showarticletitle{Optical See-through Augmented Reality Can Induce Severe Motion Sickness}.
\newblock \bibinfo{journal}{\emph{Displays}}  \bibinfo{volume}{74} (\bibinfo{date}{Sept.} \bibinfo{year}{2022}), \bibinfo{pages}{102283}.
\newblock
\showISSN{0141-9382}
\urldef\tempurl%
\url{https://doi.org/10.1016/j.displa.2022.102283}
\showDOI{\tempurl}


\bibitem[Kim et~al\mbox{.}(2018)]%
        {kimVirtualRealitySickness2018}
\bibfield{author}{\bibinfo{person}{Hyun~K. Kim}, \bibinfo{person}{Jaehyun Park}, \bibinfo{person}{Yeongcheol Choi}, {and} \bibinfo{person}{Mungyeong Choe}.} \bibinfo{year}{2018}\natexlab{}.
\newblock \showarticletitle{Virtual Reality Sickness Questionnaire ({{VRSQ}}): {{Motion}} Sickness Measurement Index in a Virtual Reality Environment}.
\newblock \bibinfo{journal}{\emph{Applied Ergonomics}}  \bibinfo{volume}{69} (\bibinfo{date}{May} \bibinfo{year}{2018}), \bibinfo{pages}{66--73}.
\newblock
\showISSN{0003-6870}
\urldef\tempurl%
\url{https://doi.org/10.1016/j.apergo.2017.12.016}
\showDOI{\tempurl}


\bibitem[Kiyokawa(2007)]%
        {kiyokawaIntroductionHeadMounted2007}
\bibfield{author}{\bibinfo{person}{Kiyoshi Kiyokawa}.} \bibinfo{year}{2007}\natexlab{}.
\newblock \showarticletitle{An {{Introduction}} to {{Head Mounted Displays}} for {{Augmented Reality}}}.
\newblock In \bibinfo{booktitle}{\emph{Emerging {{Technologies}} of {{Augmented Reality}}: {{Interfaces}} and {{Design}}}}. \bibinfo{publisher}{{IGI Global}}, \bibinfo{address}{{Osaka University, Japan}}, \bibinfo{pages}{43--63}.
\newblock
\showISBNx{978-1-59904-066-0}
\urldef\tempurl%
\url{https://doi.org/10.4018/978-1-59904-066-0.ch003}
\showDOI{\tempurl}


\bibitem[Kloft et~al\mbox{.}(2023)]%
        {kloft2023ai}
\bibfield{author}{\bibinfo{person}{Agnes~M. Kloft}, \bibinfo{person}{Robin Welsch}, \bibinfo{person}{Thomas Kosch}, {and} \bibinfo{person}{Steeven Villa}.} \bibinfo{year}{2023}\natexlab{}.
\newblock \bibinfo{title}{"AI enhances our performance, I have no doubt this one will do the same": The Placebo effect is robust to negative descriptions of AI}.
\newblock
\newblock
\showeprint[arxiv]{2309.16606}~[cs.HC]


\bibitem[Knierim et~al\mbox{.}(2020)]%
        {knierim2020demonstrating}
\bibfield{author}{\bibinfo{person}{Pascal Knierim}, \bibinfo{person}{Albrecht Schmidt}, {and} \bibinfo{person}{Thomas Kosch}.} \bibinfo{year}{2020}\natexlab{}.
\newblock \showarticletitle{{Demonstrating Thermal Flux: Using Mixed Reality to Extend Human Sight by Thermal Vision}}. In \bibinfo{booktitle}{\emph{{Proceedings of the 19th International Conference on Mobile and Ubiquitous Multimedia}}} (Duisburg-Essen, Germany) \emph{(\bibinfo{series}{MUM '20})}. \bibinfo{publisher}{ACM}, \bibinfo{address}{New York, NY, USA}, \bibinfo{pages}{348--350}.
\newblock
\urldef\tempurl%
\url{https://doi.org/10.1145/3428361.3431196}
\showDOI{\tempurl}


\bibitem[Kosch et~al\mbox{.}(2023)]%
        {koschSurveyMeasuringCognitive2023}
\bibfield{author}{\bibinfo{person}{Thomas Kosch}, \bibinfo{person}{Jakob Karolus}, \bibinfo{person}{Johannes Zagermann}, \bibinfo{person}{Harald Reiterer}, \bibinfo{person}{Albrecht Schmidt}, {and} \bibinfo{person}{Pawe{\l}~W. Wo{\'z}niak}.} \bibinfo{year}{2023}\natexlab{}.
\newblock \showarticletitle{A {{Survey}} on {{Measuring Cognitive Workload}} in {{Human-Computer Interaction}}}.
\newblock \bibinfo{journal}{\emph{Comput. Surveys}} \bibinfo{volume}{55}, \bibinfo{number}{13s} (\bibinfo{date}{July} \bibinfo{year}{2023}), \bibinfo{pages}{283:1--283:39}.
\newblock
\showISSN{0360-0300}
\urldef\tempurl%
\url{https://doi.org/10.1145/3582272}
\showDOI{\tempurl}


\bibitem[Kosch et~al\mbox{.}(2022)]%
        {koschNotiBikeAssessingTarget2022}
\bibfield{author}{\bibinfo{person}{Thomas Kosch}, \bibinfo{person}{Andrii Matviienko}, \bibinfo{person}{Florian M{\"u}ller}, \bibinfo{person}{Jessica Bersch}, \bibinfo{person}{Christopher Katins}, \bibinfo{person}{Dominik Sch{\"o}n}, {and} \bibinfo{person}{Max M{\"u}hlh{\"a}user}.} \bibinfo{year}{2022}\natexlab{}.
\newblock \showarticletitle{{{NotiBike}}: {{Assessing Target Selection Techniques}} for {{Cyclist Notifications}} in {{Augmented Reality}}}.
\newblock \bibinfo{journal}{\emph{Proceedings of the ACM on Human-Computer Interaction}} \bibinfo{volume}{6}, \bibinfo{number}{MHCI} (\bibinfo{year}{2022}), \bibinfo{pages}{1--24}.
\newblock


\bibitem[Lee and Chu(2018)]%
        {leeDualMRInteractionMixed2018}
\bibfield{author}{\bibinfo{person}{Chi-Jung Lee} {and} \bibinfo{person}{Hung-Kuo Chu}.} \bibinfo{year}{2018}\natexlab{}.
\newblock \showarticletitle{Dual-{{MR}}: Interaction with Mixed Reality Using Smartphones}. In \bibinfo{booktitle}{\emph{Proceedings of the 24th {{ACM Symposium}} on {{Virtual Reality Software}} and {{Technology}}}} \emph{(\bibinfo{series}{{{VRST}} '18})}. \bibinfo{publisher}{{Association for Computing Machinery}}, \bibinfo{address}{{New York, NY, USA}}, \bibinfo{pages}{1--2}.
\newblock
\showISBNx{978-1-4503-6086-9}
\urldef\tempurl%
\url{https://doi.org/10.1145/3281505.3281618}
\showDOI{\tempurl}


\bibitem[Liaw and Huang(2003)]%
        {liawInvestigationUserAttitudes2003}
\bibfield{author}{\bibinfo{person}{Shu-Sheng Liaw} {and} \bibinfo{person}{Hsiu-Mei Huang}.} \bibinfo{year}{2003}\natexlab{}.
\newblock \showarticletitle{An Investigation of User Attitudes toward Search Engines as an Information Retrieval Tool}.
\newblock \bibinfo{journal}{\emph{Computers in Human Behavior}} \bibinfo{volume}{19}, \bibinfo{number}{6} (\bibinfo{date}{Nov.} \bibinfo{year}{2003}), \bibinfo{pages}{751--765}.
\newblock
\showISSN{0747-5632}
\urldef\tempurl%
\url{https://doi.org/10.1016/S0747-5632(03)00009-8}
\showDOI{\tempurl}


\bibitem[McCoach et~al\mbox{.}(2013)]%
        {mccoachInstrumentDevelopmentAffective2013}
\bibfield{author}{\bibinfo{person}{D.~Betsy McCoach}, \bibinfo{person}{Robert~K. Gable}, {and} \bibinfo{person}{John~P. Madura}.} \bibinfo{year}{2013}\natexlab{}.
\newblock \bibinfo{booktitle}{\emph{Instrument {{Development}} in the {{Affective Domain}}: {{School}} and {{Corporate Applications}}}}.
\newblock \bibinfo{publisher}{{Springer}}, \bibinfo{address}{{New York, NY}}.
\newblock
\showISBNx{978-1-4614-7134-9 978-1-4614-7135-6}
\urldef\tempurl%
\url{https://doi.org/10.1007/978-1-4614-7135-6}
\showDOI{\tempurl}


\bibitem[McLean and Aldossary(2023)]%
        {mcleanDigitalTourismConsumption2023}
\bibfield{author}{\bibinfo{person}{Graeme McLean} {and} \bibinfo{person}{Mohammed Aldossary}.} \bibinfo{year}{2023}\natexlab{}.
\newblock \showarticletitle{Digital {{Tourism Consumption}}: {{The Role}} of {{Virtual Reality}} ({{VR}}) {{Vacations}} on {{Consumers}}' {{Psychological Wellbeing}}: {{An Abstract}}}. In \bibinfo{booktitle}{\emph{Optimistic {{Marketing}} in {{Challenging Times}}: {{Serving Ever-Shifting Customer Needs}}}} \emph{(\bibinfo{series}{Developments in {{Marketing Science}}: {{Proceedings}} of the {{Academy}} of {{Marketing Science}}})}, \bibfield{editor}{\bibinfo{person}{Bruna Jochims} {and} \bibinfo{person}{Juliann Allen}} (Eds.). \bibinfo{publisher}{{Springer Nature Switzerland}}, \bibinfo{address}{{Cham}}, \bibinfo{pages}{143--144}.
\newblock
\showISBNx{978-3-031-24687-6}
\urldef\tempurl%
\url{https://doi.org/10.1007/978-3-031-24687-6_53}
\showDOI{\tempurl}


\bibitem[Mehrfard et~al\mbox{.}(2019)]%
        {mehrfardComparativeAnalysisVirtual2019}
\bibfield{author}{\bibinfo{person}{Arian Mehrfard}, \bibinfo{person}{Javad Fotouhi}, \bibinfo{person}{Giacomo Taylor}, \bibinfo{person}{Tess Forster}, \bibinfo{person}{Nassir Navab}, {and} \bibinfo{person}{Bernhard Fuerst}.} \bibinfo{year}{2019}\natexlab{}.
\newblock \bibinfo{title}{A {{Comparative Analysis}} of {{Virtual Reality Head-Mounted Display Systems}}}.
\newblock
\newblock
\showeprint[arxiv]{1912.02913}~[cs]


\bibitem[Mejia and Yarosh(2017)]%
        {mejiaNineItemQuestionnaireMeasuring2017}
\bibfield{author}{\bibinfo{person}{Kenya Mejia} {and} \bibinfo{person}{Svetlana Yarosh}.} \bibinfo{year}{2017}\natexlab{}.
\newblock \showarticletitle{A {{Nine-Item Questionnaire}} for {{Measuring}} the {{Social Disfordance}} of {{Mediated Social Touch Technologies}}}.
\newblock \bibinfo{journal}{\emph{Proceedings of the ACM on Human-Computer Interaction}} \bibinfo{volume}{1}, \bibinfo{number}{CSCW} (\bibinfo{date}{Dec.} \bibinfo{year}{2017}), \bibinfo{pages}{77:1--77:17}.
\newblock
\urldef\tempurl%
\url{https://doi.org/10.1145/3134712}
\showDOI{\tempurl}


\bibitem[Moser et~al\mbox{.}(2019)]%
        {moserMixedRealityApplications2019}
\bibfield{author}{\bibinfo{person}{Thomas Moser}, \bibinfo{person}{Markus Hohlagschwandtner}, \bibinfo{person}{Gerhard {Kormann-Hainzl}}, \bibinfo{person}{Sabine P{\"o}lzlbauer}, {and} \bibinfo{person}{Josef Wolfartsberger}.} \bibinfo{year}{2019}\natexlab{}.
\newblock \showarticletitle{Mixed {{Reality Applications}} in {{Industry}}: {{Challenges}} and {{Research Areas}}}. In \bibinfo{booktitle}{\emph{Software {{Quality}}: {{The Complexity}} and {{Challenges}} of {{Software Engineering}} and {{Software Quality}} in the {{Cloud}}}} \emph{(\bibinfo{series}{Lecture {{Notes}} in {{Business Information Processing}}})}, \bibfield{editor}{\bibinfo{person}{Dietmar Winkler}, \bibinfo{person}{Stefan Biffl}, {and} \bibinfo{person}{Johannes Bergsmann}} (Eds.). \bibinfo{publisher}{{Springer International Publishing}}, \bibinfo{address}{{Cham}}, \bibinfo{pages}{95--105}.
\newblock
\showISBNx{978-3-030-05767-1}
\urldef\tempurl%
\url{https://doi.org/10.1007/978-3-030-05767-1_7}
\showDOI{\tempurl}


\bibitem[Narzt et~al\mbox{.}(2006)]%
        {narztAugmentedRealityNavigation2006}
\bibfield{author}{\bibinfo{person}{Wolfgang Narzt}, \bibinfo{person}{Gustav Pomberger}, \bibinfo{person}{Alois Ferscha}, \bibinfo{person}{Dieter Kolb}, \bibinfo{person}{Reiner M{\"u}ller}, \bibinfo{person}{Jan Wieghardt}, \bibinfo{person}{Horst H{\"o}rtner}, {and} \bibinfo{person}{Christopher Lindinger}.} \bibinfo{year}{2006}\natexlab{}.
\newblock \showarticletitle{Augmented Reality Navigation Systems}.
\newblock \bibinfo{journal}{\emph{Universal Access in the Information Society}} \bibinfo{volume}{4}, \bibinfo{number}{3} (\bibinfo{date}{March} \bibinfo{year}{2006}), \bibinfo{pages}{177--187}.
\newblock
\showISSN{1615-5297}
\urldef\tempurl%
\url{https://doi.org/10.1007/s10209-005-0017-5}
\showDOI{\tempurl}


\bibitem[Nowak et~al\mbox{.}(2020)]%
        {nowak2020what}
\bibfield{author}{\bibinfo{person}{Adam Nowak}, \bibinfo{person}{Pascal Knierim}, \bibinfo{person}{Andrzej Romanowski}, \bibinfo{person}{Albrecht Schmidt}, {and} \bibinfo{person}{Thomas Kosch}.} \bibinfo{year}{2020}\natexlab{}.
\newblock \showarticletitle{{What does the Oscilloscope Say?: Comparing the Efficiency of In-Situ Visualisations during Circuit Analysis}}. In \bibinfo{booktitle}{\emph{{Extended Abstracts of the 2020 CHI Conference on Human Factors in Computing Systems}}} (Honolulu, HI, USA) \emph{(\bibinfo{series}{CHI EA '20})}. \bibinfo{publisher}{ACM}, \bibinfo{address}{New York, NY, USA}, \bibinfo{pages}{1--7}.
\newblock
\showISBNx{978-1-4503-6819-3/20/04}
\urldef\tempurl%
\url{https://doi.org/10.1145/3334480.3382890}
\showDOI{\tempurl}


\bibitem[Numazaki et~al\mbox{.}(2017)]%
        {numazakiVREntertainmentSystem2017}
\bibfield{author}{\bibinfo{person}{Yusuke Numazaki}, \bibinfo{person}{See-Sheng Toh}, {and} \bibinfo{person}{Masanobu Endoh}.} \bibinfo{year}{Decmber 8-10, 2017}\natexlab{}.
\newblock \showarticletitle{{{VR Entertainment System}} ``{{Ideal Vacation}}'': {{A Game Designing Focused}} on the {{Sense}} of {{Presence}}}. In \bibinfo{booktitle}{\emph{The 2nd {{International Conference}} on {{Culture Technology}} ({{ICCT}})}}. \bibinfo{publisher}{{International Association for Convergence Science \& Technology}}, \bibinfo{address}{{Tokyo, Japan}}, \bibinfo{pages}{163--166}.
\newblock


\bibitem[Ram and Sheth(1989)]%
        {ramConsumerResistanceInnovations1989}
\bibfield{author}{\bibinfo{person}{S. Ram} {and} \bibinfo{person}{Jagdish~N. Sheth}.} \bibinfo{year}{1989}\natexlab{}.
\newblock \showarticletitle{Consumer {{Resistance}} to {{Innovations}}: {{The Marketing Problem}} and Its Solutions}.
\newblock \bibinfo{journal}{\emph{Journal of Consumer Marketing}} \bibinfo{volume}{6}, \bibinfo{number}{2} (\bibinfo{date}{Jan.} \bibinfo{year}{1989}), \bibinfo{pages}{5--14}.
\newblock
\showISSN{0736-3761}
\urldef\tempurl%
\url{https://doi.org/10.1108/EUM0000000002542}
\showDOI{\tempurl}


\bibitem[Raykov and Marcoulides(2011)]%
        {raykovIntroductionPsychometricTheory2011}
\bibfield{author}{\bibinfo{person}{Tenko Raykov} {and} \bibinfo{person}{George~A. Marcoulides}.} \bibinfo{year}{2011}\natexlab{}.
\newblock \bibinfo{title}{Introduction to {{Psychometric Theory}}}.
\newblock \bibinfo{howpublished}{https://www.routledge.com/Introduction-to-Psychometric-Theory/Raykov-Marcoulides/p/book/9780415878227}.
\newblock


\bibitem[Regenbrecht and Schubert(2021)]%
        {regenbrechtMeasuringPresenceAugmented2021a}
\bibfield{author}{\bibinfo{person}{Holger Regenbrecht} {and} \bibinfo{person}{Thomas Schubert}.} \bibinfo{year}{2021}\natexlab{}.
\newblock \bibinfo{title}{Measuring {{Presence}} in {{Augmented Reality Environments}}: {{Design}} and a {{First Test}} of a {{Questionnaire}}}.
\newblock
\newblock
\urldef\tempurl%
\url{https://doi.org/10.48550/arXiv.2103.02831}
\showDOI{\tempurl}
\showeprint[arxiv]{2103.02831}~[cs]


\bibitem[Rese et~al\mbox{.}(2017)]%
        {reseHowAugmentedReality2017}
\bibfield{author}{\bibinfo{person}{Alexandra Rese}, \bibinfo{person}{Daniel Baier}, \bibinfo{person}{Andreas {Geyer-Schulz}}, {and} \bibinfo{person}{Stefanie Schreiber}.} \bibinfo{year}{2017}\natexlab{}.
\newblock \showarticletitle{How Augmented Reality Apps Are Accepted by Consumers: {{A}} Comparative Analysis Using Scales and Opinions}.
\newblock \bibinfo{journal}{\emph{Technological Forecasting and Social Change}}  \bibinfo{volume}{124} (\bibinfo{date}{Nov.} \bibinfo{year}{2017}), \bibinfo{pages}{306--319}.
\newblock
\showISSN{0040-1625}
\urldef\tempurl%
\url{https://doi.org/10.1016/j.techfore.2016.10.010}
\showDOI{\tempurl}


\bibitem[Rousson et~al\mbox{.}(2002)]%
        {roussonAssessingIntraraterInterrater2002}
\bibfield{author}{\bibinfo{person}{Valentin Rousson}, \bibinfo{person}{Theo Gasser}, {and} \bibinfo{person}{Burkhardt Seifert}.} \bibinfo{year}{2002}\natexlab{}.
\newblock \showarticletitle{Assessing Intrarater, Interrater and Test{\textendash}Retest Reliability of Continuous Measurements}.
\newblock \bibinfo{journal}{\emph{Statistics in Medicine}} \bibinfo{volume}{21}, \bibinfo{number}{22} (\bibinfo{year}{2002}), \bibinfo{pages}{3431--3446}.
\newblock
\showISSN{1097-0258}
\urldef\tempurl%
\url{https://doi.org/10.1002/sim.1253}
\showDOI{\tempurl}


\bibitem[Schilit et~al\mbox{.}(1994)]%
        {4624429}
\bibfield{author}{\bibinfo{person}{B. Schilit}, \bibinfo{person}{N. Adams}, {and} \bibinfo{person}{R. Want}.} \bibinfo{year}{1994}\natexlab{}.
\newblock \showarticletitle{Context-Aware Computing Applications}. In \bibinfo{booktitle}{\emph{1994 First Workshop on Mobile Computing Systems and Applications}}. \bibinfo{publisher}{IEEE}, \bibinfo{address}{Santa Cruz, CA, USA}, \bibinfo{pages}{85--90}.
\newblock
\urldef\tempurl%
\url{https://doi.org/10.1109/WMCSA.1994.16}
\showDOI{\tempurl}


\bibitem[Schrepp et~al\mbox{.}(2017)]%
        {schreppConstructionBenchmarkUser2017a}
\bibfield{author}{\bibinfo{person}{Martin Schrepp}, \bibinfo{person}{Andreas Hinderks}, {and} \bibinfo{person}{J{\"o}rg Thomaschewski}.} \bibinfo{year}{2017}\natexlab{}.
\newblock \showarticletitle{Construction of a {{Benchmark}} for the {{User Experience Questionnaire}} ({{UEQ}})}.
\newblock \bibinfo{journal}{\emph{International Journal of Interactive Multimedia and Artificial Intelligence}} \bibinfo{volume}{4}, \bibinfo{number}{4} (\bibinfo{year}{2017}), \bibinfo{pages}{40}.
\newblock
\showISSN{1989-1660}
\urldef\tempurl%
\url{https://doi.org/10.9781/ijimai.2017.445}
\showDOI{\tempurl}


\bibitem[Schön et~al\mbox{.}(2023)]%
        {schoen2023tailor}
\bibfield{author}{\bibinfo{person}{Dominik Schön}, \bibinfo{person}{Thomas Kosch}, \bibinfo{person}{Florian Müller}, \bibinfo{person}{Martin Schmitz}, \bibinfo{person}{Sebastian Günther}, \bibinfo{person}{Lukas Bommhardt}, {and} \bibinfo{person}{Max Mühlhäuser}.} \bibinfo{year}{2023}\natexlab{}.
\newblock \showarticletitle{{Tailor Twist: Assessing Rotational Mid-Air Interactions for Augmented Reality}}. In \bibinfo{booktitle}{\emph{{Proceedings of the 2023 CHI Conference on Human Factors in Computing Systems}}} (Hamburg, Germany) \emph{(\bibinfo{series}{CHI '23})}. \bibinfo{publisher}{ACM}, \bibinfo{address}{New York, NY, USA}, \bibinfo{pages}{1--14}.
\newblock
\showISBNx{978-1-4503-9421-5/23/04}
\urldef\tempurl%
\url{https://doi.org/10.1145/3544548.3581461}
\showDOI{\tempurl}


\bibitem[Shrestha(2021)]%
        {shresthaFactorAnalysisTool2021}
\bibfield{author}{\bibinfo{person}{Noora Shrestha}.} \bibinfo{year}{2021}\natexlab{}.
\newblock \showarticletitle{Factor {{Analysis}} as a {{Tool}} for {{Survey Analysis}}}.
\newblock \bibinfo{journal}{\emph{American Journal of Applied Mathematics and Statistics}} \bibinfo{volume}{9}, \bibinfo{number}{1} (\bibinfo{date}{Jan.} \bibinfo{year}{2021}), \bibinfo{pages}{4--11}.
\newblock
\showISSN{2328-7306}
\urldef\tempurl%
\url{https://doi.org/10.12691/ajams-9-1-2}
\showDOI{\tempurl}


\bibitem[Slater et~al\mbox{.}(2020)]%
        {slaterEthicsRealismVirtual2020}
\bibfield{author}{\bibinfo{person}{Mel Slater}, \bibinfo{person}{Cristina {Gonzalez-Liencres}}, \bibinfo{person}{Patrick Haggard}, \bibinfo{person}{Charlotte Vinkers}, \bibinfo{person}{Rebecca {Gregory-Clarke}}, \bibinfo{person}{Steve Jelley}, \bibinfo{person}{Zillah Watson}, \bibinfo{person}{Graham Breen}, \bibinfo{person}{Raz Schwarz}, \bibinfo{person}{William Steptoe}, \bibinfo{person}{Dalila Szostak}, \bibinfo{person}{Shivashankar Halan}, \bibinfo{person}{Deborah Fox}, {and} \bibinfo{person}{Jeremy Silver}.} \bibinfo{year}{2020}\natexlab{}.
\newblock \showarticletitle{The {{Ethics}} of {{Realism}} in {{Virtual}} and {{Augmented Reality}}}.
\newblock \bibinfo{journal}{\emph{Frontiers in Virtual Reality}}  \bibinfo{volume}{1} (\bibinfo{date}{March} \bibinfo{year}{2020}), \bibinfo{pages}{1}.
\newblock
\showISSN{2673-4192}
\urldef\tempurl%
\url{https://doi.org/10.3389/frvir.2020.00001}
\showDOI{\tempurl}


\bibitem[Speicher et~al\mbox{.}(2019)]%
        {speicherWhatMixedReality2019}
\bibfield{author}{\bibinfo{person}{Maximilian Speicher}, \bibinfo{person}{Brian~D. Hall}, {and} \bibinfo{person}{Michael Nebeling}.} \bibinfo{year}{2019}\natexlab{}.
\newblock \showarticletitle{What Is {{Mixed Reality}}?}. In \bibinfo{booktitle}{\emph{Proceedings of the 2019 {{CHI Conference}} on {{Human Factors}} in {{Computing Systems}}}} \emph{(\bibinfo{series}{{{CHI}} '19})}. \bibinfo{publisher}{{Association for Computing Machinery}}, \bibinfo{address}{{Glasgow, Scotland Uk}}, \bibinfo{pages}{1--15}.
\newblock
\urldef\tempurl%
\url{https://doi.org/10.1145/3290605.3300767}
\showDOI{\tempurl}


\bibitem[Thomas and Holmquist(2021)]%
        {thomasFunctionalityAllThat2021}
\bibfield{author}{\bibinfo{person}{Derianna Thomas} {and} \bibinfo{person}{Lars~Erik Holmquist}.} \bibinfo{year}{2021}\natexlab{}.
\newblock \showarticletitle{Is {{Functionality All That Matters}}? {{Examining Everyday User Opinions}} of {{Augmented Reality Devices}}}. In \bibinfo{booktitle}{\emph{2021 {{IEEE Conference}} on {{Virtual Reality}} and {{3D User Interfaces Abstracts}} and {{Workshops}} ({{VRW}})}}. \bibinfo{publisher}{{IEEE}}, \bibinfo{address}{{Lisbon, Portugal}}, \bibinfo{pages}{232--237}.
\newblock
\urldef\tempurl%
\url{https://doi.org/10.1109/VRW52623.2021.00050}
\showDOI{\tempurl}


\bibitem[Venkatesh and Bala(2008)]%
        {venkateshTechnologyAcceptanceModel2008}
\bibfield{author}{\bibinfo{person}{Viswanath Venkatesh} {and} \bibinfo{person}{Hillol Bala}.} \bibinfo{year}{2008}\natexlab{}.
\newblock \showarticletitle{Technology {{Acceptance Model}} 3 and a {{Research Agenda}} on {{Interventions}}}.
\newblock \bibinfo{journal}{\emph{Decision Sciences}} \bibinfo{volume}{39}, \bibinfo{number}{2} (\bibinfo{year}{2008}), \bibinfo{pages}{273--315}.
\newblock
\showISSN{1540-5915}
\urldef\tempurl%
\url{https://doi.org/10.1111/j.1540-5915.2008.00192.x}
\showDOI{\tempurl}


\bibitem[Venkatesh and Davis(2000)]%
        {venkateshTheoreticalExtensionTechnology2000}
\bibfield{author}{\bibinfo{person}{Viswanath Venkatesh} {and} \bibinfo{person}{Fred~D. Davis}.} \bibinfo{year}{2000}\natexlab{}.
\newblock \bibinfo{title}{A {{Theoretical Extension}} of the {{Technology Acceptance Model}}: {{Four Longitudinal Field Studies}} {\textbar} {{Management Science}}}.
\newblock \bibinfo{howpublished}{https://pubsonline.informs.org/doi/abs/10.1287/mnsc.46.2.186.11926}.
\newblock


\bibitem[Villa et~al\mbox{.}(2023a)]%
        {villa2023the}
\bibfield{author}{\bibinfo{person}{Steeven Villa}, \bibinfo{person}{Thomas Kosch}, \bibinfo{person}{Felix Grelka}, \bibinfo{person}{Albrecht Schmidt}, {and} \bibinfo{person}{Robin Welsch}.} \bibinfo{year}{2023}\natexlab{a}.
\newblock \showarticletitle{{The Placebo Effect of Human Augmentation: Anticipating Cognitive Augmentation Increases Risk-Taking Behavior}}.
\newblock \bibinfo{journal}{\emph{{Computers in Human Behavior}}} \bibinfo{volume}{146}, \bibinfo{number}{C} (\bibinfo{year}{2023}), \bibinfo{pages}{107787}.
\newblock
\showISSN{0747-5632}
\urldef\tempurl%
\url{https://doi.org/10.1016/j.chb.2023.107787}
\showDOI{\tempurl}


\bibitem[Villa et~al\mbox{.}(2023b)]%
        {villaSocietyAttitudesHuman2022}
\bibfield{author}{\bibinfo{person}{Steeven Villa}, \bibinfo{person}{Jasmin Niess}, \bibinfo{person}{Albrecht Schmidt}, {and} \bibinfo{person}{Robin Welsch}.} \bibinfo{year}{2023}\natexlab{b}.
\newblock \showarticletitle{Society's Attitudes Towards Human Augmentation and Performance Enhancement Technologies (SHAPE) Scale}.
\newblock \bibinfo{journal}{\emph{Proceedings of the ACM on Interactive, Mobile, Wearable and Ubiquitous Technologies}} \bibinfo{volume}{7}, \bibinfo{number}{3} (\bibinfo{year}{2023}), \bibinfo{pages}{1--23}.
\newblock


\bibitem[Wilson and McGill(2018)]%
        {wilsonViolentVideoGames2018}
\bibfield{author}{\bibinfo{person}{Graham Wilson} {and} \bibinfo{person}{Mark McGill}.} \bibinfo{year}{2018}\natexlab{}.
\newblock \showarticletitle{Violent {{Video Games}} in {{Virtual Reality}}: {{Re-Evaluating}} the {{Impact}} and {{Rating}} of {{Interactive Experiences}}}. In \bibinfo{booktitle}{\emph{Proceedings of the 2018 {{Annual Symposium}} on {{Computer-Human Interaction}} in {{Play}}}} \emph{(\bibinfo{series}{{{CHI PLAY}} '18})}. \bibinfo{publisher}{{Association for Computing Machinery}}, \bibinfo{address}{{New York, NY, USA}}, \bibinfo{pages}{535--548}.
\newblock
\showISBNx{978-1-4503-5624-4}
\urldef\tempurl%
\url{https://doi.org/10.1145/3242671.3242684}
\showDOI{\tempurl}


\bibitem[Witmer and Singer(1998)]%
        {witmerMeasuringPresenceVirtual1998}
\bibfield{author}{\bibinfo{person}{Bob~G. Witmer} {and} \bibinfo{person}{Michael~J. Singer}.} \bibinfo{year}{1998}\natexlab{}.
\newblock \showarticletitle{Measuring {{Presence}} in {{Virtual Environments}}: {{A Presence Questionnaire}}}.
\newblock \bibinfo{journal}{\emph{Presence: Teleoperators and Virtual Environments}} \bibinfo{volume}{7}, \bibinfo{number}{3} (\bibinfo{date}{June} \bibinfo{year}{1998}), \bibinfo{pages}{225--240}.
\newblock
\urldef\tempurl%
\url{https://doi.org/10.1162/105474698565686}
\showDOI{\tempurl}


\bibitem[Wo{\'z}niak et~al\mbox{.}(2021)]%
        {wozniakCreepyTechnologyWhat2021}
\bibfield{author}{\bibinfo{person}{Pawe{\l}~W. Wo{\'z}niak}, \bibinfo{person}{Jakob Karolus}, \bibinfo{person}{Florian Lang}, \bibinfo{person}{Caroline Eckerth}, \bibinfo{person}{Johannes Sch{\"o}ning}, \bibinfo{person}{Yvonne Rogers}, {and} \bibinfo{person}{Jasmin Niess}.} \bibinfo{year}{2021}\natexlab{}.
\newblock \showarticletitle{Creepy {{Technology}}:{{What Is It}} and {{How Do You Measure It}}?}. In \bibinfo{booktitle}{\emph{Proceedings of the 2021 {{CHI Conference}} on {{Human Factors}} in {{Computing Systems}}}}. \bibinfo{publisher}{{ACM}}, \bibinfo{address}{{Yokohama Japan}}, \bibinfo{pages}{1--13}.
\newblock
\showISBNx{978-1-4503-8096-6}
\urldef\tempurl%
\url{https://doi.org/10.1145/3411764.3445299}
\showDOI{\tempurl}


\bibitem[Zhang and Preacher(2015)]%
        {zhangFactorRotationStandard2015}
\bibfield{author}{\bibinfo{person}{Guangjian Zhang} {and} \bibinfo{person}{Kristopher~J. Preacher}.} \bibinfo{year}{2015}\natexlab{}.
\newblock \bibinfo{title}{Factor {{Rotation}} and {{Standard Errors}} in {{Exploratory Factor Analysis}} - {{Guangjian Zhang}}, {{Kristopher J}}. {{Preacher}}, 2015}.
\newblock \bibinfo{howpublished}{https://journals.sagepub.com/doi/10.3102/1076998615606098}.
\newblock


\end{thebibliography}

\end{document}